\documentclass[]{elsart}

\usepackage[bookmarks]{hyperref}
\usepackage{subfigure}
\usepackage{graphicx}
\usepackage{threeparttablex}
\usepackage{lscape}
\usepackage{epsfig}
\usepackage{rotating}
\usepackage{ifthen}
\usepackage{multicol}

\usepackage{amssymb}
\usepackage{amsmath}
\usepackage{xspace}
\usepackage{epstopdf}
\usepackage{mathtools}
\usepackage{longtable}
\usepackage[title]{appendix}

\newcommand{\xe}{\ensuremath{\rm ^{136}Xe}\xspace}
\newcommand{\KTHU}{K, Th and U\xspace}
\newcommand{\hnot}{\ensuremath{\rm HNO_3}\xspace}

\newcommand{\ife}[3]{\ifthenelse{\equal{#1}{#2}}{#3}{{}}}
\newcommand{\inch}{\ensuremath{^{\prime\prime}}\xspace}
\newcommand{\ppq}{\ensuremath{\rm fg/g}\xspace}
\newcommand{\ppt}{\ensuremath{\rm pg/g}\xspace}
\newcommand{\ppb}{\ensuremath{\rm ng/g}\xspace}
\newcommand{\ppm}{\ensuremath{\rm \mu g/g}\xspace}
\newcommand{\idx}[1]{%
\ife{#1}{331}	{226}%
\ife{#1}{332}	{227}%
\ife{#1}{333}	{228}%
\ife{#1}{334}	{229}%
\ife{#1}{460}	{230}%
\ife{#1}{}	{}%
\ife{#1}{}	{}%
\ife{#1}{322}	{231}%
\ife{#1}{323}	{232}%
\ife{#1}{324}	{233}%
\ife{#1}{338}	{234}%
\ife{#1}{339}	{235}%
\ife{#1}{340}	{236}%
\ife{#1}{341}	{237}%
\ife{#1}{342}	{238}%
\ife{#1}{347}	{239}%
\ife{#1}{344}	{240}%
\ife{#1}{345}	{241}%
\ife{#1}{312}	{242}%
\ife{#1}{354}	{243}%
\ife{#1}{391}	{244}%
\ife{#1}{392}	{245}%
\ife{#1}{393}	{246}%
\ife{#1}{394}	{247}%
\ife{#1}{335}	{248}%
\ife{#1}{336}	{249}%
\ife{#1}{337}	{250}%
\ife{#1}{367}	{251}%
\ife{#1}{368}	{252}%
\ife{#1}{}	{}%
\ife{#1}{}	{}%
\ife{#1}{361}	{253}%
\ife{#1}{475}	{254}%
\ife{#1}{476}	{255}%
\ife{#1}{477}	{256}%
\ife{#1}{478}	{257}%
\ife{#1}{479}	{258}%
\ife{#1}{375}	{259}%
\ife{#1}{376}	{260}%
\ife{#1}{377}	{261}%
\ife{#1}{362}	{262}%
\ife{#1}{365}	{263}%
\ife{#1}{366}	{264}%
\ife{#1}{}	{}%
\ife{#1}{}	{}%
\ife{#1}{303}	{265}%
\ife{#1}{359}	{266}%
\ife{#1}{307}	{267}%
\ife{#1}{308}	{268}%
\ife{#1}{309}	{269}%
\ife{#1}{310}	{270}%
\ife{#1}{314}	{271}%
\ife{#1}{315}	{272}%
\ife{#1}{317}	{273}%
\ife{#1}{321}	{274}%
\ife{#1}{355}	{275}%
\ife{#1}{356}	{276}%
\ife{#1}{357}	{277}%
\ife{#1}{319}	{278}%
\ife{#1}{320}	{279}%
\ife{#1}{442}	{280}%
\ife{#1}{398}	{281}%
\ife{#1}{399}	{282}%
\ife{#1}{400}	{283}%
\ife{#1}{401}	{284}%
\ife{#1}{402}	{285}%
\ife{#1}{443}	{286}%
\ife{#1}{384}	{287}%
\ife{#1}{378}	{288}%
\ife{#1}{484}	{289}%
\ife{#1}{379}	{290}%
\ife{#1}{485}	{291}%
\ife{#1}{380}	{292}%
\ife{#1}{486}	{293}%
\ife{#1}{385}	{294}%
\ife{#1}{386}	{295}%
\ife{#1}{318}	{296}%
\ife{#1}{418}	{297}%
\ife{#1}{482}	{298}%
\ife{#1}{417}	{299}%
\ife{#1}{419}	{300}%
\ife{#1}{473}	{301}%
\ife{#1}{422}	{302}%
\ife{#1}{370}	{303}%
\ife{#1}{483}	{304}%
\ife{#1}{436}	{305}%
\ife{#1}{404}	{306}%
\ife{#1}{437}	{307}%
\ife{#1}{487}	{308}%
\ife{#1}{488}	{309}%
\ife{#1}{428}	{310}%
\ife{#1}{429}	{311}%
\ife{#1}{447}	{312}%
\ife{#1}{465}	{313}%
\ife{#1}{466}	{314}%
\ife{#1}{467}	{315}%
\ife{#1}{481}	{316}%
\ife{#1}{}	{}%
\ife{#1}{}	{}%
\ife{#1}{}	{}%
\ife{#1}{}	{}%
\ife{#1}{}	{}%
\ife{#1}{}	{}%
\ife{#1}{}	{}%
\ife{#1}{}	{}%
\ife{#1}{}	{}%
\ife{#1}{}	{}%
\ife{#1}{}	{}%
\ife{#1}{}	{}%
\ife{#1}{}	{}%
\ife{#1}{}	{}%
\ife{#1}{}	{}%
\ife{#1}{}	{}%
\ife{#1}{}	{}%
\ife{#1}{}	{}%
\ife{#1}{}	{}%
\ife{#1}{}	{}%
\ife{#1}{}	{}%
\ife{#1}{}	{}%
\ife{#1}{}	{}%
\ife{#1}{}	{}%
\ife{#1}{}	{}%
\ife{#1}{}	{}%
\ife{#1}{}	{}%
\ife{#1}{}	{}%
\ife{#1}{}	{}%
\ife{#1}{}	{}%
\ife{#1}{}	{}%
\ife{#1}{}	{}%
\ife{#1}{}	{}%
\ife{#1}{}	{}%
\ife{#1}{}	{}%
\ife{#1}{}	{}%
\ife{#1}{}	{}%
\ife{#1}{}	{}%
\ife{#1}{}	{}%
\ife{#1}{}	{}%
\ife{#1}{}	{}%
\ife{#1}{}	{}%
\ife{#1}{}	{}%
\ife{#1}{}	{}%
\ife{#1}{1}	{1}%
\ife{#1}{2}	{2}%
\ife{#1}{3}	{3}%
\ife{#1}{4}	{4}%
\ife{#1}{5}	{5}%
\ife{#1}{6}	{6}%
\ife{#1}{7}	{7}%
\ife{#1}{8}	{8}%
\ife{#1}{9}	{9}%
\ife{#1}{10}	{10}%
\ife{#1}{11}	{11}%
\ife{#1}{12}	{12}%
\ife{#1}{13}	{13}%
\ife{#1}{14}	{14}%
\ife{#1}{15}	{15}%
\ife{#1}{16}	{16}%
\ife{#1}{17}	{17}%
\ife{#1}{18}	{18}%
\ife{#1}{19}	{19}%
\ife{#1}{20}	{20}%
\ife{#1}{21}	{21}%
\ife{#1}{22}	{22}%
\ife{#1}{23}	{23}%
\ife{#1}{24}	{24}%
\ife{#1}{25}	{25}%
\ife{#1}{26}	{26}%
\ife{#1}{27}	{27}%
\ife{#1}{28}	{28}%
\ife{#1}{29}	{29}%
\ife{#1}{30}	{30}%
\ife{#1}{31}	{31}%
\ife{#1}{32}	{32}%
\ife{#1}{33}	{33}%
\ife{#1}{34}	{34}%
\ife{#1}{35}	{35}%
\ife{#1}{36}	{36}%
\ife{#1}{37}	{37}%
\ife{#1}{38}	{38}%
\ife{#1}{39}	{39}%
\ife{#1}{40}	{40}%
\ife{#1}{41}	{41}%
\ife{#1}{42}	{42}%
\ife{#1}{43}	{43}%
\ife{#1}{44}	{44}%
\ife{#1}{45}	{45}%
\ife{#1}{46}	{46}%
\ife{#1}{47}	{47}%
\ife{#1}{48}	{48}%
\ife{#1}{49}	{49}%
\ife{#1}{50}	{50}%
\ife{#1}{51}	{51}%
\ife{#1}{52}	{52}%
\ife{#1}{53}	{53}%
\ife{#1}{54}	{54}%
\ife{#1}{55}	{55}%
\ife{#1}{56}	{56}%
\ife{#1}{57}	{57}%
\ife{#1}{58}	{58}%
\ife{#1}{59}	{59}%
\ife{#1}{60}	{60}%
\ife{#1}{61}	{61}%
\ife{#1}{62}	{62}%
\ife{#1}{63}	{63}%
\ife{#1}{64}	{64}%
\ife{#1}{65}	{65}%
\ife{#1}{66}	{66}%
\ife{#1}{67}	{67}%
\ife{#1}{68}	{68}%
\ife{#1}{69}	{69}%
\ife{#1}{70}	{70}%
\ife{#1}{71}	{71}%
\ife{#1}{72}	{72}%
\ife{#1}{73}	{73}%
\ife{#1}{74}	{74}%
\ife{#1}{75}	{75}%
\ife{#1}{76}	{76}%
\ife{#1}{77}	{77}%
\ife{#1}{78}	{78}%
\ife{#1}{79}	{79}%
\ife{#1}{80}	{80}%
\ife{#1}{81}	{81}%
\ife{#1}{82}	{82}%
\ife{#1}{83}	{83}%
\ife{#1}{84}	{84}%
\ife{#1}{85}	{85}%
\ife{#1}{86}	{86}%
\ife{#1}{87}	{87}%
\ife{#1}{88}	{88}%
\ife{#1}{89}	{89}%
\ife{#1}{90}	{90}%
\ife{#1}{91}	{91}%
\ife{#1}{92}	{92}%
\ife{#1}{93}	{93}%
\ife{#1}{94}	{94}%
\ife{#1}{95}	{95}%
\ife{#1}{96}	{96}%
\ife{#1}{97}	{97}%
\ife{#1}{98}	{98}%
\ife{#1}{99}	{99}%
\ife{#1}{100}	{100}%
\ife{#1}{101}	{101}%
\ife{#1}{102}	{102}%
\ife{#1}{103}	{103}%
\ife{#1}{104}	{104}%
\ife{#1}{105}	{105}%
\ife{#1}{106}	{106}%
\ife{#1}{107}	{107}%
\ife{#1}{108}	{108}%
\ife{#1}{109}	{109}%
\ife{#1}{110}	{110}%
\ife{#1}{111}	{111}%
\ife{#1}{112}	{112}%
\ife{#1}{113}	{113}%
\ife{#1}{114}	{114}%
\ife{#1}{115}	{115}%
\ife{#1}{116}	{116}%
\ife{#1}{117}	{117}%
\ife{#1}{118}	{118}%
\ife{#1}{119}	{119}%
\ife{#1}{120}	{120}%
\ife{#1}{121}	{121}%
\ife{#1}{122}	{122}%
\ife{#1}{123}	{123}%
\ife{#1}{124}	{124}%
\ife{#1}{125}	{125}%
\ife{#1}{126}	{126}%
\ife{#1}{127}	{127}%
\ife{#1}{128}	{128}%
\ife{#1}{129}	{129}%
\ife{#1}{130}	{130}%
\ife{#1}{131}	{131}%
\ife{#1}{132}	{132}%
\ife{#1}{133}	{133}%
\ife{#1}{134}	{134}%
\ife{#1}{135}	{135}%
\ife{#1}{136}	{136}%
\ife{#1}{137}	{137}%
\ife{#1}{138}	{138}%
\ife{#1}{139}	{139}%
\ife{#1}{140}	{140}%
\ife{#1}{141}	{141}%
\ife{#1}{142}	{142}%
\ife{#1}{143}	{143}%
\ife{#1}{144}	{144}%
\ife{#1}{145}	{145}%
\ife{#1}{146}	{146}%
\ife{#1}{147}	{147}%
\ife{#1}{148}	{148}%
\ife{#1}{149}	{149}%
\ife{#1}{150}	{150}%
\ife{#1}{151}	{151}%
\ife{#1}{152}	{152}%
\ife{#1}{153}	{153}%
\ife{#1}{154}	{154}%
\ife{#1}{155}	{155}%
\ife{#1}{156}	{156}%
\ife{#1}{157}	{157}%
\ife{#1}{158}	{158}%
\ife{#1}{159}	{159}%
\ife{#1}{160}	{160}%
\ife{#1}{161}	{161}%
\ife{#1}{162}	{162}%
\ife{#1}{163}	{163}%
\ife{#1}{164}	{164}%
\ife{#1}{165}	{165}%
\ife{#1}{166}	{166}%
\ife{#1}{167}	{167}%
\ife{#1}{168}	{168}%
\ife{#1}{169}	{169}%
\ife{#1}{170}	{170}%
\ife{#1}{171}	{171}%
\ife{#1}{172}	{172}%
\ife{#1}{173}	{173}%
\ife{#1}{174}	{174}%
\ife{#1}{175}	{175}%
\ife{#1}{176}	{176}%
\ife{#1}{177}	{177}%
\ife{#1}{178}	{178}%
\ife{#1}{179}	{179}%
\ife{#1}{180}	{180}%
\ife{#1}{181}	{181}%
\ife{#1}{182}	{182}%
\ife{#1}{183}	{183}%
\ife{#1}{184}	{184}%
\ife{#1}{185}	{185}%
\ife{#1}{186}	{186}%
\ife{#1}{187}	{187}%
\ife{#1}{188}	{188}%
\ife{#1}{189}	{189}%
\ife{#1}{190}	{190}%
\ife{#1}{191}	{191}%
\ife{#1}{192}	{192}%
\ife{#1}{193}	{193}%
\ife{#1}{194}	{194}%
\ife{#1}{195}	{195}%
\ife{#1}{196}	{196}%
\ife{#1}{197}	{197}%
\ife{#1}{198}	{198}%
\ife{#1}{199}	{199}%
\ife{#1}{200}	{200}%
\ife{#1}{201}	{201}%
\ife{#1}{202}	{202}%
\ife{#1}{203}	{203}%
\ife{#1}{204}	{204}%
\ife{#1}{205}	{205}%
\ife{#1}{206}	{206}%
\ife{#1}{207}	{207}%
\ife{#1}{208}	{208}%
\ife{#1}{209}	{209}%
\ife{#1}{210}	{210}%
\ife{#1}{211}	{211}%
\ife{#1}{212}	{212}%
\ife{#1}{213}	{213}%
\ife{#1}{214}	{214}%
\ife{#1}{215}	{215}%
\ife{#1}{216}	{216}%
\ife{#1}{217}	{217}%
\ife{#1}{218}	{218}%
\ife{#1}{219}	{219}%
\ife{#1}{220}	{220}%
\ife{#1}{221}	{221}%
\ife{#1}{222}	{222}%
\ife{#1}{223}	{223}%
\ife{#1}{224}	{224}%
\ife{#1}{225}	{225}%
}

\begin{document}

\newenvironment{cfigure}[1][tbp]{\begin{figure}[#1]\centering}{\end{figure}}
\newenvironment{cfigure1c}[1][tbp]{\begin{figure*}[#1]\centering}{\end{figure*}}


\newcommand\geniso[2]{\ensuremath{\rm ^{#2}#1}\xspace}

\newcommand\newiso[3][]{
\ifx&#1&
  \expandafter\newcommand\csname #2\endcsname[1][#3]{\geniso{#2}{##1}}
\else
  \expandafter\newcommand\csname #1\endcsname[1][#3]{\geniso{#2}{##1}}
\fi
}


%
%

\newiso{K}{40}
\newiso{U}{238}
\newiso{Th}{232}
\newiso{Co}{60}
\newiso{Ra}{}
\newiso{Rn}{}
\newiso{Bi}{}
\newiso{Pb}{210}
\newiso{Tl}{208}
\newiso{Ac}{}
\newiso{W}{}
\newiso{Hf}{}
\newiso{Cs}{}
\newiso{Ca}{}
\newiso{Sb}{}
\newiso{La}{}
\newiso{Cr}{}
\newiso{Zn}{}
\newiso{As}{}
\newiso{Au}{}
\newiso{Ti}{}
\newiso{Sc}{}
\newiso{Xe}{}
\newiso{Ar}{40}
\newiso{Na}{22}
\newiso{I}{125}
\newiso{Te}{}
\newiso{Po}{}
\newiso{Am}{241}
\newiso{Fe}{55}
\newiso{Mo}{100}
\newiso{Zr}{}
\newiso{Nd}{}
\newiso{Tc}{}
\newiso{Ba}{}
\newiso{Y}{}
\newiso{Np}{}

\begin{frontmatter}

\title{Trace radioactive impurities in final construction materials for EXO-200}

\author[IBS]{D.S.~Leonard},
\author[Alabama]{D.~Auty\thanksref{Auty}}\thanks[Auty]{Now at University of Alberta, Edmonton, Canada},
\author[Alabama]{T.~Didberidze},
\author[Carleton,TRIUMF]{R.~Gornea},
\author[INMS]{P. Grinberg},
\author[SDakota]{R.~MacLellan},
\author[INMS]{B. Methven},
\author[Alabama]{A.~Piepke},
\author[Bern]{J.-L.~Vuilleumier},
\author[Indiana]{J.B.~Albert},
% ...

\address[IBS]{IBS Center for Underground Physics, Daejeon 34047, Korea}
\address[Alabama]{Department of Physics and Astronomy, University of Alabama, Tuscaloosa, Alabama 35487, USA}
\address[Carleton]{Physics Department, Carleton University, Ottawa, Ontario K1S 5B6, Canada}
\address[TRIUMF]{TRIUMF, Vancouver, British Columbia V6T 2A3, Canada}
\address[INMS]{Measurement Science and Standards, National Research Council Canada, Ottawa ON, Canada}
\address[SDakota]{Physics Department, University of South Dakota, Vermillion, South Dakota 57069, USA}
\address[Bern]{LHEP, Albert Einstein Center, University of Bern, Bern, Switzerland}
\address[Indiana]{Physics Department and CEEM, Indiana University, Bloomington, Indiana 47405, USA}
% ...

\begin{abstract}
We report results from a systematic measurement campaign conducted to identify low radioactivity materials for the construction of the EXO-200 double beta decay experiment. Partial results from this campaign have already been reported in a 2008 paper by the EXO collaboration. Here we release the remaining data, collected since 2007, to the public. The data reported were obtained using a variety of analytic techniques.  The measurement sensitivities are among the best in the field. Construction of the EXO-200 detector has been concluded, and Phase-I data was taken from 2011 to 2014. The detector's extremely low background implicitly verifies the measurements and the analysis assumptions made during construction and reported in this paper.


\end{abstract}

\end{frontmatter}


\section{Introduction}
\label{sec:Intro}

Low energy, low-rate counting experiments such as searches for double beta decay, dark matter, and neutrino oscillations rely on access to construction materials containing the smallest possible amounts of radioactivity. The presence of radioactivity near the detector, even in ultra-trace concentrations, often causes unwanted background, potentially limiting the scientific reach of these experiments. The access to a range of low activity materials is, therefore, enabling science. 

Specifically, this work was motivated by the Enriched Xenon Observatory (EXO), a multi-stage experimental research program with the purpose of detecting rare double beta decays of \xe~\cite{EXO}. With EXO-200, we search for these decays in an underground cryogenic time-projection chamber (TPC) filled with approximately 110~kg of active liquid xenon enriched to 80\% in \xe. In Ref.~\cite{rad1} we reported on a campaign of measurements of radioactive impurities in potential construction materials for the purpose of achieving the low background rates required for successful operation. Similar measurement campaigns have been published for rare-event search efforts~\cite{Arpesella02,Laubenstein04,Budjas09,Aprile11,Armengaud13,Alvarez13,Abgrall16,Wang16}. Here we augment the previously reported measurements with results obtained during the final stages of design and construction of EXO-200.  Measurement techniques and conditions were generally the same as those described in Ref.~\cite{rad1}.  As in the previous work, the radio-assay campaign described here focuses on natural radioactivity, namely \K[40], \Th and \U.

EXO-200 started taking data in 2011. The experiment has been described in detail in~\cite{EXOpartI}. The experiment performed the first observation of the two-neutrino double beta decay of \Xe[136]~\cite{exo_2nbb_2011}, placed stringent limits on the neutrinoless decay mode~\cite{exo_0nbb_2012,exo_nature}, and reported the most precise determination of any two-neutrino double beta decay rate~\cite{exo_2nbb_2014}. The background event rate of $R_b=(1.7\pm 0.2)\cdot 10^{-3}$~keV$^{-1}$kg$^{-1}$yr$^{-1}$~\cite{exo_0nbb_2012,exo_2nbb_2014} observed with the EXO-200 detector, around  the double beta decay Q-value of $Q_{\beta\beta}=2457.83 \pm 0.37$ keV~\cite{qvalue}, is one of the lowest in its field. A detailed background analysis has been published in Ref.~\cite{exo_bkg_2015}. This analysis compared the data-derived estimates of the activity contents of detector components with those obtained in the radio-assay program. In an alternate approach the radio-assay values were fed into the detector simulation to arrive at expectation values. Detector background predictions which were made before data-taking agree reasonably well with the observed rate. It was further noted that for most components the radio-assay program yielded stronger constraints than the data driven analysis~\cite{exo_bkg_2015}.  The EXO-200 detector thus provides some validation of the data, methods and assumptions reported in this work and the previous EXO-200 component radioactivity compilation~\cite{rad1}. 

The EXO-200 materials analysis effort was structured around a detailed, GEANT~3.21 based Monte Carlo simulation of the experiment. A total background budget of 33 events per year in 110 kg of xenon (after cuts) for events in the energy interval $Q_{\beta\beta}\pm 2\sigma_{\beta\beta}$ was defined~\cite{EXOpartI}, where $\sigma_{\beta\beta}$ stands for the energy resolution at the Q-value. A target value of  $\sigma_{\beta\beta}/Q_{\beta\beta}=0.015$ was chosen. Major experiment components, such as the cryostat or the lead shield, were allowed to contribute 10\% of the total budget while small components were given a 1\% background allowance. This fuzzy scheme allowed material acceptance decisions to be made before all components had been specified and analyzed for their radioactivity content. The background allowance was then translated into a maximally allowable radioactivity content for each component by means of the Monte Carlo model. This allowance determined, in turn, the choice of analysis method.  All materials and components used during the EXO-200 construction were subject to this process; no exceptions were made. 

The results of the EXO-200 radioactivity screening program are reported as element concentrations, in units of g/g (grams of impurity per gram of sample).  Multiplication with conversion factors of $3.17\cdot 10^4$~(Bq/kg)/(g/g) (\K[40]), $4.07\cdot 10^6$~(Bq/kg)/(g/g) (\Th) and $1.23\cdot 10^7$~(Bq/kg)/(g/g) (\U) yields nuclide specific activities in units of Bq/kg. 

The following analysis methods, described below, were employed in this work and in the previous measurement campaign:

\begin{enumerate}
\item Above-ground and below-ground low-background gamma-ray spectroscopy using Ge detectors.
\label{ge_counting}
\item Glow-discharge mass spectroscopy (GDMS) \label{GDMS} 
\item Inductively Coupled Plasma Mass Spectroscopy (ICPMS) \label{ICPMS} 
\item Neutron Activation Analysis (NAA) \label{NAA}
\end{enumerate}

The radioactivity-induced background for the EXO-200 search for neutrinoless double beta decay stems, to a large extent, from \Tl[208] (Th-series) and \Bi[214] (U-series) $\beta+\gamma$ decays. \Co[60] coincidence summing and cosmogenic activation play a small role too. 
Germanium counting has the lowest analysis sensitivity but determines the relevant \Tl[208] and \Bi[214] decay rates directly. No further assumptions are needed to convert radioactivity concentrations into background event rates. 
GDMS, ICPMS and NAA offer improved sensitivity; however, they determine the concentrations of the long-lived heads of the decay series, \Th and \U, respectively. The Th and U decay rates can only be related to those of \Tl[208] and \Bi[214] by making assumptions about the establishment of chain equilibrium. For the EXO-200 preparation, for results reported in Ref.~\cite{rad1}, and for results reported here, it was decided to assume that the decay chains are in secular equilibrium. The fact that background estimates made prior to data taking agree reasonably well with the EXO-200 observation ultimately justified this assumption. Improved analysis sensitivity therefore comes at the expense of reliance on assumptions. In a sense, one risk has to be balanced against another. 

\section{Underground Gamma Counting}

All of the gamma counting was done at the Vue-des-Alpes laboratory, located in a lateral cavern in the Vue-des-Alpes road tunnel, in the Swiss Jura. Some measurements were carried out above ground with two Ge detectors used also for NAA. They are described in Section~\ref{sec:gamfit}. The vertical rock overburden is 230~m, corresponding to 600~mwe (see Refs.~\cite{rad1,Gonin03}). The rate of cosmic muons through a flat horizontal surface is around 0.2~m$^{-2}$s$^{-1}$, about a factor 1\,000 less than above ground. Cosmic-ray neutrons are completely eliminated. A Ge detector made by Eurisys (now Canberra) in 2001 was used. It is a p-type coaxial device. The volume of the germanium crystal is 400~cm$^3$. The active volume is reduced by a dead layer of a few times 100~$\rm \mu$m on the outer side and the top. The dead layer on the inner hole is negligible. The crystal is housed in a can-shaped vessel forming the end of a cryostat and made from ultra-clean P\'{e}chiney aluminum with relatively low Z. The end cap is particularly thin (0.5~mm), providing a good $\gamma$ transmission even at low energy. The germanium crystal is mounted on a copper cold finger, cooled by liquid nitrogen. 

The detector is surrounded by a shield made from electrolytic tough pitch (ETP) copper inside, with a  thickness varying from 12.5~cm to 20~cm, and by 15~cm to 20~cm of lead outside.  This efficiently suppresses local $\gamma$ activities. The shielding rests on a steel table. The liquid nitrogen dewar is located outside, and the cold finger traverses the shielding. The shielding copper was taken over and adapted from a Ge experiment in the Gothard tunnel~\cite{Reusser92}.  The lead was purchased new.  Both the copper and lead were originally obtained from local companies and samples of both had been tested with an older Ge detector with a sensitivity of order 10~\ppb to U and Th.

The radon level in the lab was measured to fluctuate around 85~Bq/m$^3$. To suppress the background from \Rn[222], the shielding itself is enclosed in an aluminum box resting on the  steel table. The box is airtight. It is over-pressurized with boil-off nitrogen from an external dewar, also used to refill the germanium dewar. The top of the aluminium box and the top of the shielding can be opened with a hoist to insert samples into a cavity surrounding the detector. 

The energy resolution is 1.4~keV FWHM at 238~keV, and 2.5~keV at 1460~keV, scaling with the the square root of the energy above that. At low energy the response is nearly Gaussian, while a low-energy tail appears at higher energies. 

\begin{figure}[th]
	\begin{center}
		\includegraphics[width=10.0cm]{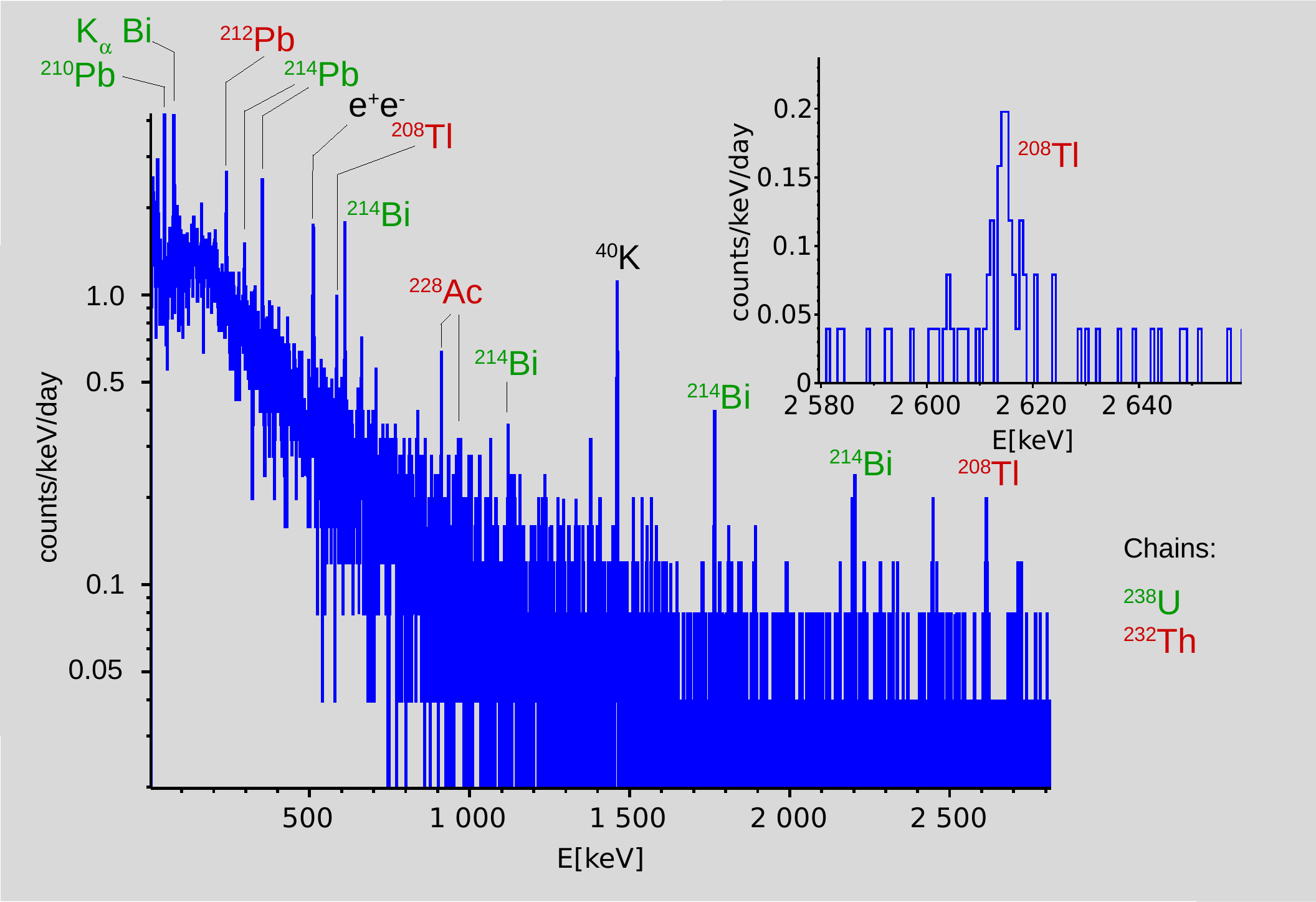}
	\end{center}
		\caption{The background spectrum of the Ge counter at the Vue-des-Alpes laboratory, accumulated during 882.5 h. \label{fig:linesUThK}}
		\label{fi:bkgd}
\end{figure}

\subsection{Standard operations} Depending on their nature, samples were cleaned before testing. Metallic samples were wiped with acetone and cleaned in an ultrasonic bath with methanol only. Samples with plastic pieces were wiped with methanol. Samples were usually packed in plastic bags or plastic vials to avoid contaminating the shielding. This work was done in a clean room. Samples were inserted into the cavity surrounding the detector, where they could be placed in a geometry optimizing the acceptance. The first 24 hours of data after closing the shielding was discarded to make sure that all the radon contamination had been purged out. Measurements usually lasted one week, but were extended to up to four weeks depending on the activity found and the required sensitivity. 	 

\subsection{Count rates}
\begin{table}[bt]
	\begin{center}
		\begin{tabular}{l|r| r | l | c}
			chain      &    nuclide   & E$_{\gamma}$(keV) & \hspace{0.3cm}BR & Background [d$^{-1}$]  \\				
			\hline 
\U              & \Pb[214]   &    295.20         & 0.185    & 1.91  $\pm$ 0.33 \\
			& \Pb[214]   &    351.92         & 0.358    & 3.25  $\pm$ 0.39 \\
			& \Bi[214]   &    609.30         & 0.448    & 2.87  $\pm$ 0.34 \\
			& \Bi[214]   &   1120.28         & 0.148    & 0.64  $\pm$ 0.17 \\
			& \Bi[214]   &   1764.50         & 0.154    & 0.69  $\pm$ 0.16 \\
			\hline                      
			& \Pb[210]   &     46.54         & 0.0425   & 4.17  $\pm$ 0.47 \\ 
			\hline                        
\Th              & \Ac[228]   &    338.30         & 0.113    & 0.45  $\pm$ 0.25 \\
			& \Ac[228]   &    911.20         & 0.266    & 0.66  $\pm$ 0.19 \\
			& \Ac[228]   &    968.97         & 0.162    & 0.34  $\pm$ 0.14 \\
			& \Pb[212]   &    238.63         & 0.433    & 4.07  $\pm$ 0.46 \\
			& \Tl[208]   &    583.19         & 0.3055   & 1.13  $\pm$ 0.24 \\ 
			& \Tl[208]   &   2614.53         & 0.3585   & 0.72  $\pm$ 0.15 \\
			\hline                        
			& \K[40]     &   1460.86         & 0.1066   & 3.00  $\pm$ 0.30 \\
			\hline                      
			& \Cs[137]   &    661.60         & 0.851    & 0.99  $\pm$ 0.23 \\
			\hline
			& \Co[60]    &   1173.23         & 0.9999   & 0.07  $\pm$ 0.12 \\
			& \Co[60]    &   1332.50         & 0.9998   & 0.11  $\pm$ 0.11 \\  
			
		\end{tabular}
	\end{center}
	\caption{List of the lines with energy E$_{\gamma}$ and branching ratio BR~\cite{Nudat2} considered in the routine analysis of samples. 
	The nuclides within a chain are arranged in chronological order. The internal background of the Vue-des-Alpes detector is also given. \label{ta:linesUThK}}
\end{table}

The count rates of the most relevant lines were obtained by integrating the counts in a window with a width of 2 FWHM. The background under the line was extrapolated from a window 4 to 8 times as wide and was subtracted. 

The lines with energy {E$_{\gamma}$ and branching ratio BR \cite{Nudat2} given in table \ref{ta:linesUThK} were analyzed routinely. They include the main lines in the \U and \Th chains, the lines resulting from \K[40], \Cs[137], and \Co[60] decays. 

\subsection{The internal background} The background from internal activities, or originating from the shielding, was measured at regular intervals without any sample, and was observed to be fairly constant. The integral rate from 5 keV to 2.8 MeV was 26.6$\pm$0.17~h$^{-1}$. The spectrum is displayed in Fig. \ref{fi:bkgd}. The \K[40] line dominates, but weak lines from the \U, \Th chains are also visible. The count rates are listed in Table \ref{ta:linesUThK}. 

The continuum is largely due to Compton scattering accompanying the observed $\gamma$ transitions, but also in part to direct cosmic $\mu$ hits and to  bremsstrahlung generated by these $\mu$'s in the lead and copper shielding.
The background spectrum was analyzed with the same procedure as that for the samples. The rates obtained for each of the lines listed in Table \ref{ta:linesUThK} was taken as background, and subtracted from the rates with a given sample, yielding the final count rate attributed to that sample. All uncertainties were added in quadrature. 

\subsection{Evaluation of the activity} The inspection of the integral count rate from 5~keV to 2.8~MeV provides a first check of the radiopurity of a sample. Any excess above the background value indicates a contamination. The analysis of the transitions mentioned above allowed more specificity.  The detector acceptance as a function of energy was calculated with GEANT3 for each sample geometry and chemical composition. 

For \K[40] and \Cs[137] only one transition is available, and the activity is obviously that obtained from the counts in the corresponding peak. For \Co[60], the average from the two transitions at 1173 and 1332~keV was computed. For the \U and \Th chains, if the data were consistent with secular equilibrium, the weighted average of all the transitions in Table \ref{ta:linesUThK} was calculated. One exception was the 46.5 keV transition in \Pb[210] in the \U chain, which was not included. It gives in general a comparatively poor precision, because of low acceptance. Sometimes the secular equilibrium appears broken at that level. \Pb[210] is listed separately. 

If an activity was observed above background at the 90~\% confidence level for any of the mentioned nuclides, the nuclide was declared active. In all other cases an upper limit was calculated using the renormalized Gaussian method, which takes into account that the true activity cannot be negative.  

The probability distribution for the true activity is taken as a Gaussian function centered at the measured activity. The statistical uncertainty at the 68~\% confidence level of the measured activity was taken as the standard deviation. Only the parameter space corresponding to true activities larger or equal to zero is retained, and the Gaussian function is normalized to 1 in that parameter space. To calculate the limit, the probability distribution is integrated from zero upward, until the fractional area corresponding to the desired confidence level is reached.

The quoted activities for \U and \Th are effective activities, assuming secular equilibrium. The real values may differ, but the effective activities are a precise measure of the penetrating $\gamma$ backgrounds, the most worrisome type of backgrounds for materials outside of the sensitive medium itself. The $\alpha$ and $\beta$ emissions from outer materials are mostly blocked from reaching the sensitive volume. 

\section{GDMS}

Glow Discharge Mass Spectrometry (GDMS) is one of the most comprehensive trace element analysis techniques currently available for the direct determination of the composition of conductive solid materials. This is especially important for high purity materials where low detection limits are desirable. Since samples are analyzed in solid form, laborious and error-prone dissolution procedures inherent to such techniques as Inductively Coupled Plasma Mass Spectrometry are avoided.

In GDMS, the sample functions as the cathode, whereas the reaction cell forms the anode portion of this two electrode system. A low flow of pure argon sustains a DC electrical discharge at low pressure in which sputtering of the sample occurs.  Application of several hundred volts between the electrodes establishes the discharge, producing a low-pressure plasma containing electrons and Ar ions.  The major voltage drop occurs close to the sample cathode and leads to the sputtering of the surface by bombardment with energetic Ar ions.  Atoms sputtered from the surface enter the plasma wherein they are rapidly ionized by a number of processes, including collisions with energetic electrons as well as Penning ionization.  One of the principal advantages of the GD technique is that it allows the determination of the bulk composition of the sample, assuming intrinsic homogeneity.  The ionized atoms are then extracted into the mass spectrometer for separation based upon their mass/charge ratio followed by detection. 

Under the typical discharge conditions, with the Thermo Fisher VG 9000 GDMS instrument used for our work,  the surface of the sample is ablated at a rate of roughly 1~$\rm \mu m/min$. It should also be noted that all samples are chemically pre-cleaned prior to analysis to remove any surface contamination that may have occurred during the cutting/shaping processing of a sample into a pin-shaped geometry suitable for the instrument. The sample is then sputtered for up to 30 minutes prior to acquisition of analytical data, to ensure that all possible surface contamination has been removed.  Analyte concentrations are obtained as the ratio of the ion current from each impurity detected in the sample to that from the sample matrix (the ``matrix'' refers to the primary or overall composition of the sample, as opposed to individual trace impurities).  These data are then corrected for minor element dependent changes in relative sputter yields using relative sensitivity factors (RSF).  The latter are determined by pre-calibrating the instrument using reference material of known elemental composition.  Results are typically valid within a factor of two for all elements except C, N and O, for which a factor of five is typical.  However, these elements are not relevant for the current work.

A unique feature of GDMS is that these RSF values, although instrument specific, do not change over time, are stable for years and matrix effects are low enough such that semi-quantitative, order-of-magnitude or better measurements can be made even without matrix-matched calibration standards.

The high sensitivity of this analytical technique enables determination of impurities in the sample at the order of 10~pg/g if using a sufficient analysis time (high number of mass scans).  Also, the universality of the method allows for the detection of almost all elements in the periodic table in conductive inorganic matrices.

A potential drawback of GDMS is that it is a micro-sampling technique requiring a solid sample and, as such, is subject to the homogeneity of that portion of the sample sputtered into the ionization volume. As sample inhomogeneity cannot be accounted for, it may contribute to a substantial increase in the reported estimated uncertainties. As there are very few reference materials available for high purity metals and semiconductor materials, the uncertainty associated with the RSF correction factors can be significant. These facts inherently limit the accuracy and relegate the method, for our purposes, to a semi-quantitative technique.

\section{ICPMS}
Inductively Coupled Plasma Mass Spectrometry (ICPMS) is one of the most powerful analytical methods for trace and ultra-trace analysis, offering sub \ppt detection limits for U and Th with minimal analysis time. Samples, most commonly in the form of liquid solutions, are introduced into an argon gas stream.  The resulting aerosol mixture is ionized in an RF field, creating a plasma.  Molecules are dissociated, and most elements are ionized with high efficiency.  Ions are extracted electrostatically into a mass spectrometer.  The technique offers higher sensitivity than GDMS, and the liquid sample form allows for easy preparation of calibrated standards or spiked sample solutions.  This allows calibrations to be performed in identical conditions to the sample measurements, thus providing highly quantitative results. 

The clear difficulty for ICPMS is the need to provide the sample in a suitable liquid form, typically aqueous.  This can sometimes require development of specialized dissolution techniques using strong, high-purity acids and, for more difficult samples, specialized equipment as well such as a microwave digestion system, asher, etc.  Metals are typically the easiest samples to dissolve in acids, but unlike for GDMS, non-conductive samples can be analyzed if suitable dissolution techniques exist. While the instruments can measure concentrations in liquid from a few~\ppq up to hundreds of~\ppm, the sensitivity of the technique is generally limited by the need to keep the total dissolved solid concentration low, typically far below~1\%.  High matrix levels give rise to deposition of matrix constituents on the surface of the interface cones (metal orifices where ions pass from high pressure plasma to low pressure spectrometer), causing significant signal drift.  Once the samples are in a liquid form, it is also possible to enhance sensitivities by developing techniques to pre-concentrate the elements of interest relative to the matrix.  Analyte separation and pre-concentration has become of paramount importance in order to achieve sub-\ppt level analysis.  However, all chemicals used in sample processing must be carefully controlled to avoid introduction of background contaminants. Analysis blanks must be subtracted, producing systematic uncertainties which are often a limiting factor for improvements in sensitivity.

For the work reported here we followed sample digestion by separation of analytes from the matrix.  This can be achieved by passing the digested sample through a chemical filter which can separate the analytes from the matrix. Specifically we used UTEVA resins from Eichrom Industries, Inc. consisting of diamyl amylphosphonate (DAAP) extractant adsorbed onto an inert polyacrylamide support consisting of particles having an external diameter of 100- 150~$\rm \mu$m~\cite{Gri05} and packed into chromatographic columns.  The analytes are initially captured by the extractant while the matrix passes through. Uranium can than be eluted from the column with 0.02~M HCl, and 0.5\% oxalic acid was required and used in order to elute Th. Both fractions were evaporated to near dryness; the Th fraction was further decomposed with a 1:1 mixture of concentrated HNO$_3$ and 30\% H$_2$O$_2$, once again evaporated to near dryness and reconstituted to 2~ml with 0.5\% nitric acid. This extra step was necessary to decompose organic extractants from the column to achieve a stable detection efficiency, thereby further elevating the blank. 

All steps of this procedure need to be monitored closely in order to ensure accurate results.  Purity of the chemicals, materials and reagents used in the analysis must also be verified. Acids used for the dissolution and separation procedures were of super-pure analytical grade (TAMA PureAA-100 or higher) All lab-ware used was cleaned with nitric acid for at least 24~h. Sample digestion and cleaning procedures were conducted in a class 100 clean room. Separations were undertaken in a class 10 fume hood. Blanks included all reagents and were run in parallel with samples. New resins exhibited Th concentrations of approximately 8~\ppt.  After thorough cleaning with several alternate washings with HCl, oxalic acid and double-distilled water (DDW), thorium concentration levels were reduced to about 1.5 \ppt. 
The recovery efficiencies for U and Th were approximately 80\% and 60\% respectively.   A typical initial sample mass of 1~g represented by a 2~ml reconstituted volume results in an effective sample-to-solution ratios of roughly 35\% by mass.  In comparison, a direct digestion measurement without the separation step is typically performed at concentrations of around 0.1 \% dissolved solid, and typically, for the Perkin Elmer Elan-DRC II used for this work,  can achieve limits of detection (LODs) for U and Th of about 15~ppt in a solid sample.  The pre-concentration step represents about a 350 times signal enhancement, potentially resulting in LOD's as low as 45~\ppq.  The actual LOD's achieved are however significantly limited by the previously mentioned backgrounds arising from the resins and reagents, and to some extent by the small reconstituted solution volume. We note that since these data were taken, further development of this basic method has led to detection limits at the level of 10~\ppq, achieved by improving the background levels~\cite{PNNLCu}.

Procedural LODs of about 0.5 and 1.5~\ppt in solids for U and Th respectively were achieved, based on the dissolution of 1~g samples and final reconstitution to 2~ml. Prior to digestion of a sample, a surface cleaning procedure was performed in order to remove any impurities present on the surface of the sample. This was achieved by placing the sample into acid solutions for a period of time. The acid and strength used, as well as the time period, depended on the matrix of the sample. For example, surface cleaning for Cu matrix samples were performed by placing the sample into a 3M~\hnot solution for 20~minutes; aqua regia was used for Al.
  
Thorium showed a severe memory effect during the determination by ICPMS as it tends to adhere to the walls of the pump tubing, spray chamber and cones, making longer washing times necessary between samples or standards. Also, low recoveries for Th were observed when using a Pyrex beaker during the evaporation step. This was overcome by using Teflon beakers instead.

\section{NAA}
\label{sec:NAA}
Neutron activation analysis (NAA) is a well-established trace element analysis technique. It is based on the capture of thermal and epi-thermal neutrons by a stable or meta-stable nuclear species. In case the newly created nuclear species has a shorter half-life than the mother, the decay rate is boosted. This works if a sufficient fraction of the target nuclei is transmuted. Depending on the capture cross section (and thus nuclear species), this requires the use of high neutron-flux research reactors to achieve an enhancement in the decay rate.  It is applicable to materials where the matrix, does not form long lived radioactivity after neutron capture. Many of the analysis details were discussed in our previous paper~\cite{rad1} and will not be repeated. We will, instead, provide a more detailed discussion of the procedures used to derive elemental concentrations from measured $\gamma$-ray spectra. 

The activations reported here utilized the MIT Reactor (MITR) as neutron source. Sample preparation and counting were done at EXO labs. MITR is a 6 MW tank-type reactor, utilizing highly enriched uranium fuel, resulting in a compact fuel assembly. It has a light-water moderator and heavy-water and graphite neutron reflectors. The activations were performed at reactor powers varying between 4.8 and 5.8 MW. MITR has two pneumatic sample delivery systems. One with 2'' diameter (2PH1), offering high thermal and fast neutron fluxes (up to $\rm 5.5\cdot 10^{13}\; cm^{-2}s^{-1}$ observed in our studies). A second tube with 1'' diameter (1PH1) shows lower thermal neutron flux and a much reduced fast neutron flux. We utilized 1PH1 for samples with low metal content to reduce background caused by fast-neutron induced side reactions. All work presented here is based on irradiations in 2PH1. Both insertion tubes are spatially located in the reflector region of the reactor. The neutron spectrum is similar to that of a light water reactor. Its parametrization is discussed later in this paper.

NAA usually utilizes radiative neutron capture on some target element of interest $T$, forming the unstable daughter species $D$: $T(n,\gamma)D$. For activation products with sufficiently long life times, $\tau_D$, the delayed beta decay of $D$ can be measured, for example by utilizing resultant $\gamma$-radiation and high resolution Ge-detectors. The irradiation of a sample, containing $N_T$ atoms of the species of interest and lasting the time $t_i$, results in a time $t$  dependent nuclear decay rate $R_D$ of the activation product: 
\begin{equation}
R_D(t)\; =\; N_T \cdot \left(1-e^{-t_i/\tau_D}\right)
\cdot e^{-t/\tau_D}\cdot \int_{0}^{\infty} \Phi(E) \cdot \sigma(E)\; dE, \label{eq:daughter_decay_rate}
\end{equation}
where $\Phi(E)$ denotes the energy dependent neutron flux the sample is exposed to while inside the reactor, $\sigma(E)$ is the differential neutron capture cross section, and $t=0$ is chosen as the end of the irradiation. Equation~\ref{eq:daughter_decay_rate} holds for irradiations consuming only a small fraction of the target atoms. Equation~\ref{eq:daughter_decay_rate} can be used to derive $N_T$ (the quantity of interest) from a measured decay rate of the activation daughter.    This requires calculation of the reaction yields per atom given by the integral in Equation~\ref{eq:daughter_decay_rate}, and of the time corrections (the exponential factors in Equation~\ref{eq:daughter_decay_rate}). Appendix~\ref{sec:neutrons} of this article describes the procedures used to determine the yields from tabulated cross sections and measured neutron fluxes. 

Application of this analysis tool allowed to routinely reach Th/U sensitivities of {\it \ppt} level or bettter. 

\subsection{Sample preparation and recovery}

Both sample preparation and recovery followed a well-defined protocol. This is important for achieving reproducible results. NAA results are typically reported as element concentrations per unit sample mass. This normalization assumes that the impurities are dominantly located in the sample bulk. Special attention therefore needs to be given to the removal of any surface impurities before sample analysis. Before activation the samples were handled in a class 500 clean room. The sample surfaces were cleaned with DI water, organic solvents and dilute HNO$_3$, all analytical grade or better. All beakers used in this process were pre-cleaned using the same sequence of solvents. After the cleaning and for activations of less than 10~h duration, the samples were packed in 5~ml polyethylene (PE) bottles (cleaned in 5~ml solvent and acid) that were then sealed using a soldering iron. Leak tightness of the containers was verified in a heated water bath. For long term activations samples were sealed into cleaned quartz containers. All samples and their containers were carefully weighed using an analytic balance to provide the normalization and to quantify the post-activation sample recovery fraction. The fact that sub-\ppt sensitivities are routinely reached, even for small samples, verifies the validity of these cleaning procedures. 

After activation, the handling focused on the clean separation of the sample from the, often externally contaminated, sample bottle. A separate glove box was used. PE bottles were softened during the activation. This limits the activation time at MITR to about 10~h, after which these bottles tend to disintegrate, leading to sample loss. Sample removal was achieved by carefully cutting off the top of the bottle. The sample was then removed by pouring it into a clean PE counting bottle or retrieving it with tweezers. Care was needed so as not to transfer any of the abundant contamination on the outside of the counting bottle onto the sample. All beakers and instruments were only used once to avoid any carry-over problems. After recovery the samples were weighed.
 
Recovering activated samples from quartz vials requires a different procedure. Because of their mechanical strength, quartz vials need to be cracked inside a metal tube (to prevent explosive pressure relief). As this procedure does not allow to separate the sample from the bottle's external surfaces, the bottles needed to be vigorously cleaned with solvents and acids before cracking. The cleaning agents were counted after each cleaning, the bottles were cracked after no more activity was detected.
 
Depending on the activated sample, recovery can hindered if the sample decomposes during activation. In this case, sometimes only a fraction of the sample can be recovered cleanly and safely.

\subsection{Gamma-ray fitting}
\label{sec:gamfit}
The activated samples are counted on three high resolution Ge detectors, commencing on average 33 hours after the end of irradiation. Gamma-ray spectra are accumulated initially over 1~h time intervals. As time goes on and the short-lived activities die out, longer and longer accumulation times are chosen to obtain reasonable statistics in each run. This allows following the decay of the unstable nuclides. A measurement of the half-life constitutes an additional observable. Gamma-ray peak energies $E_{\gamma}$ determine the decaying parent nuclide $j$. Fitting peak $k$, the integral $I(E_{\gamma,k},t_i)$, determined at time $t_i$ (combined with gamma energy dependent counting efficiency $\varepsilon(E_{\gamma,k})$ branching ratio $b_{k,j}$ and run counting time $t_{ci}$) provides an estimate of the activity $A_{k,j}(t_i)$ of the parent nuclide and its error $\sigma_{A_{k,j}}(t_i)$: 
\begin{equation}
A_{k,j}(t_i)\; =\; \frac{I(E_{\gamma,k},t_i)}{t_{ci}\cdot \varepsilon(E_{\gamma,k})\cdot b_{k,j}}
\end{equation}

A custom peak fitting code called GDFIT performs a coupled fit to all $\gamma$ lines associated with a particular nuclide and thus summarizes them into one nuclide-specific activity.  More detail is provided in Ref.~\cite{rad1}. The $\chi^2$-minimizations are based on the CERN software package MINUIT. This procedure is essentially equivalent to performing a weighted average of all single gamma-line activities, minimizing the counting error. $\varepsilon(E_{\gamma,k})$ is determined using a calibrated mixed nuclide solution for various counting geometries. Analysis consistency is routinely verified by counting activated samples on all three of the available detectors and comparing the derived sample activities.

\begin{figure}[htb!!!]
\centering
\includegraphics[width=120mm]{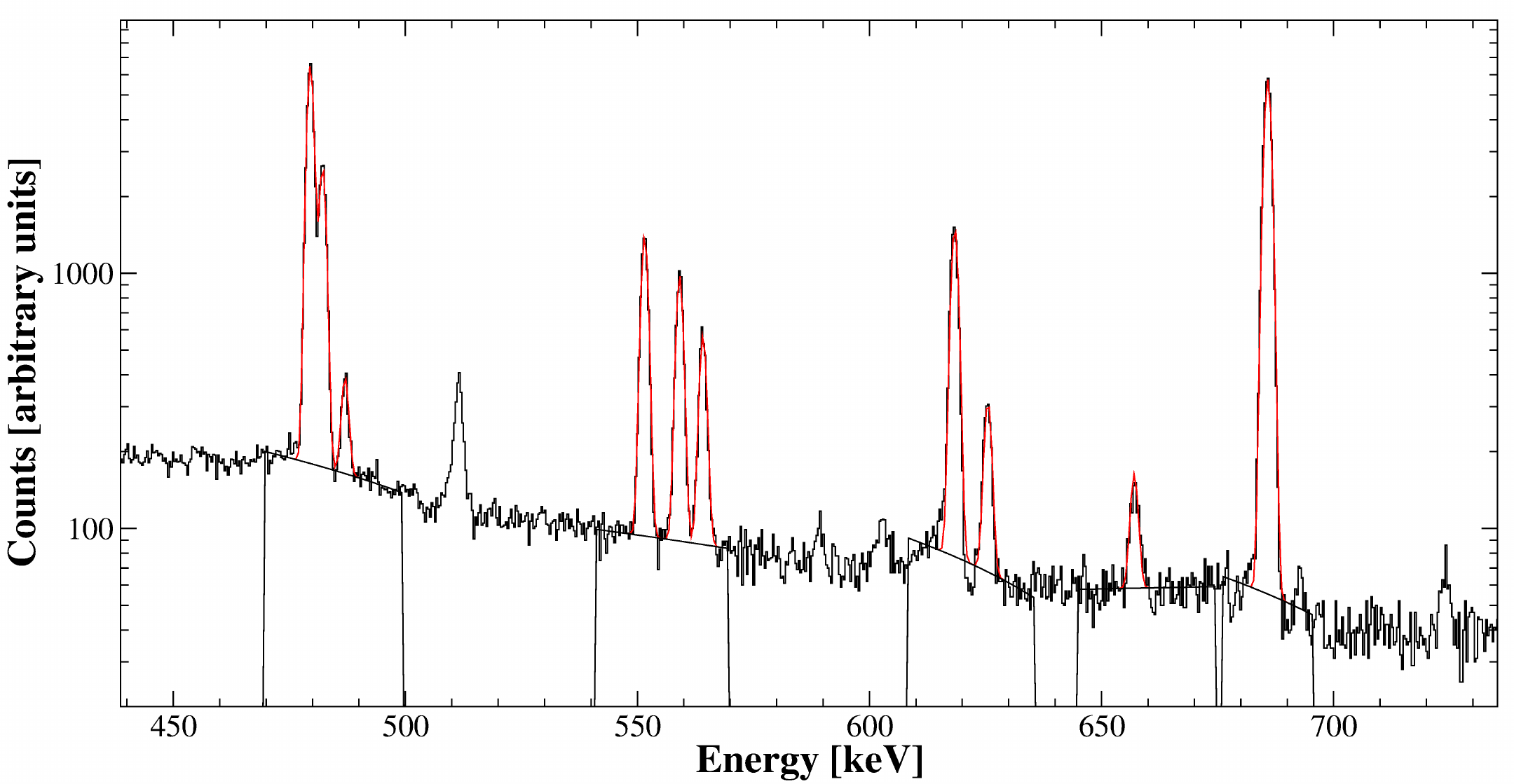}
\caption{\label{fig:g_spect}Example for a simultaneous gamma line fit performed for the
480, 552, 618, 626 and 686~keV lines of \W[187], the 559 and 657~keV lines
of \As[76] and the 482, 489 and 564~keV lines of \Hf[181], \Ca[47], \Sb[122],
respectively.}
\end{figure}
Because of the multitude of peaks present in any of the spectra, care has to be taken to include in the fit gamma lines of similar energy to, but not associated with, the activity of interest.  These can produce peak area degeneracy and/or can influence the background fit if not properly included.  For multiple-gamma activities, constraints through the linked  branching ratios can help significantly to resolve such degeneracies.  If the interfering decay also has multiple peaks then  even degeneracy of perfectly overlapping peaks can be resolved while still providing added statistical benefit for the activity of interest.  Interferences that are not properly fit can often be spotted as a deviation of the half-life from its expectation value.  In the worst case scenario, degeneracy can be resolved by fitting multiple half-life components to the time series.  In principle this could be included as part of a single global fit to peaks at all time intervals, but this was not implemented and could have practical challenges.

All gamma peaks of interest are fitted in all $\gamma$-spectra and then collected as a function of time. The various reported nuclide activities are plotted as a function of time, as shown in Figure~\ref{fig:np239_act}.
\begin{figure}[htb!!!]
\subfigure[]{\includegraphics[width=\textwidth/2]{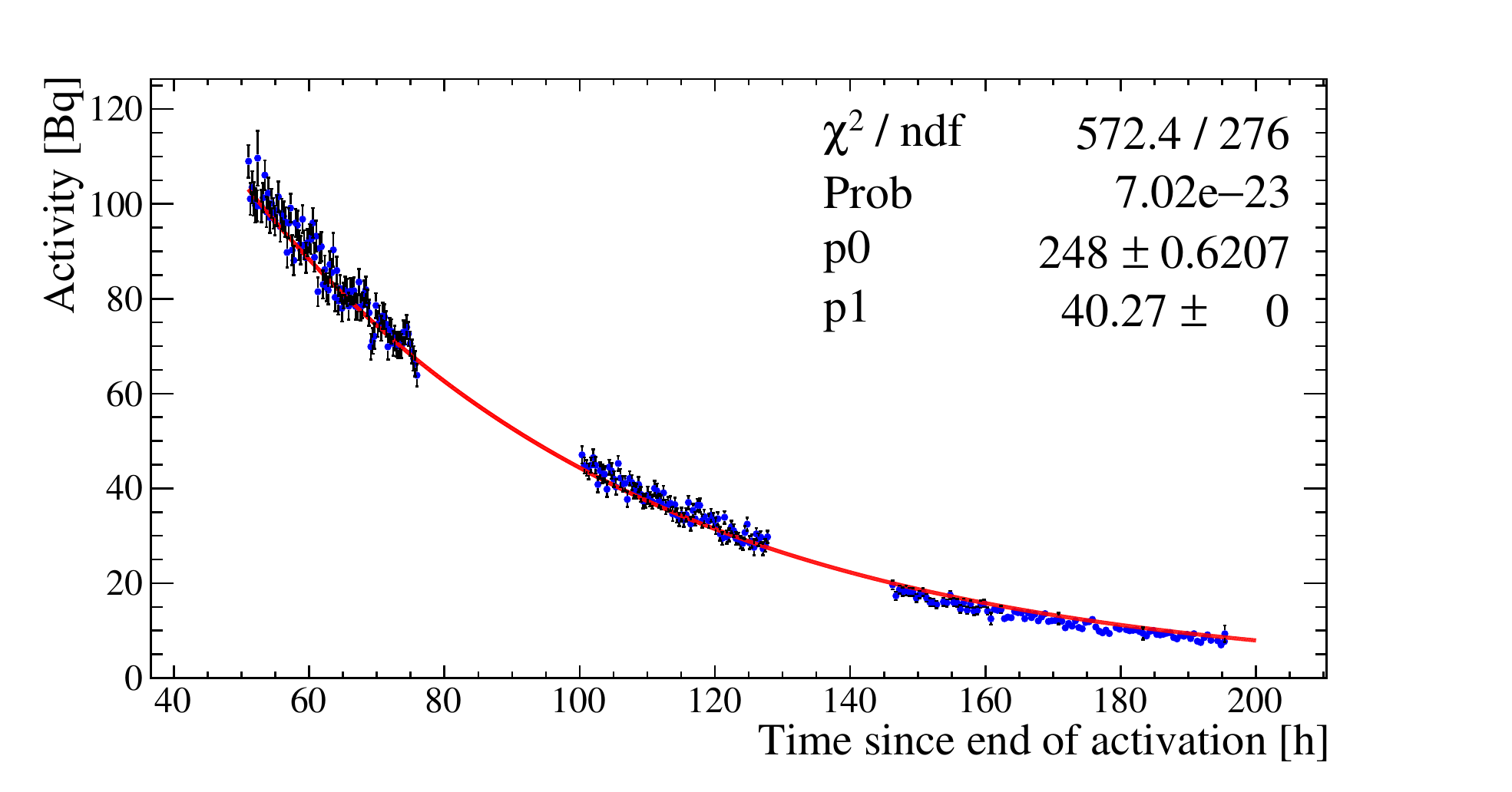}}
\subfigure[]{\includegraphics[width=\textwidth/2]{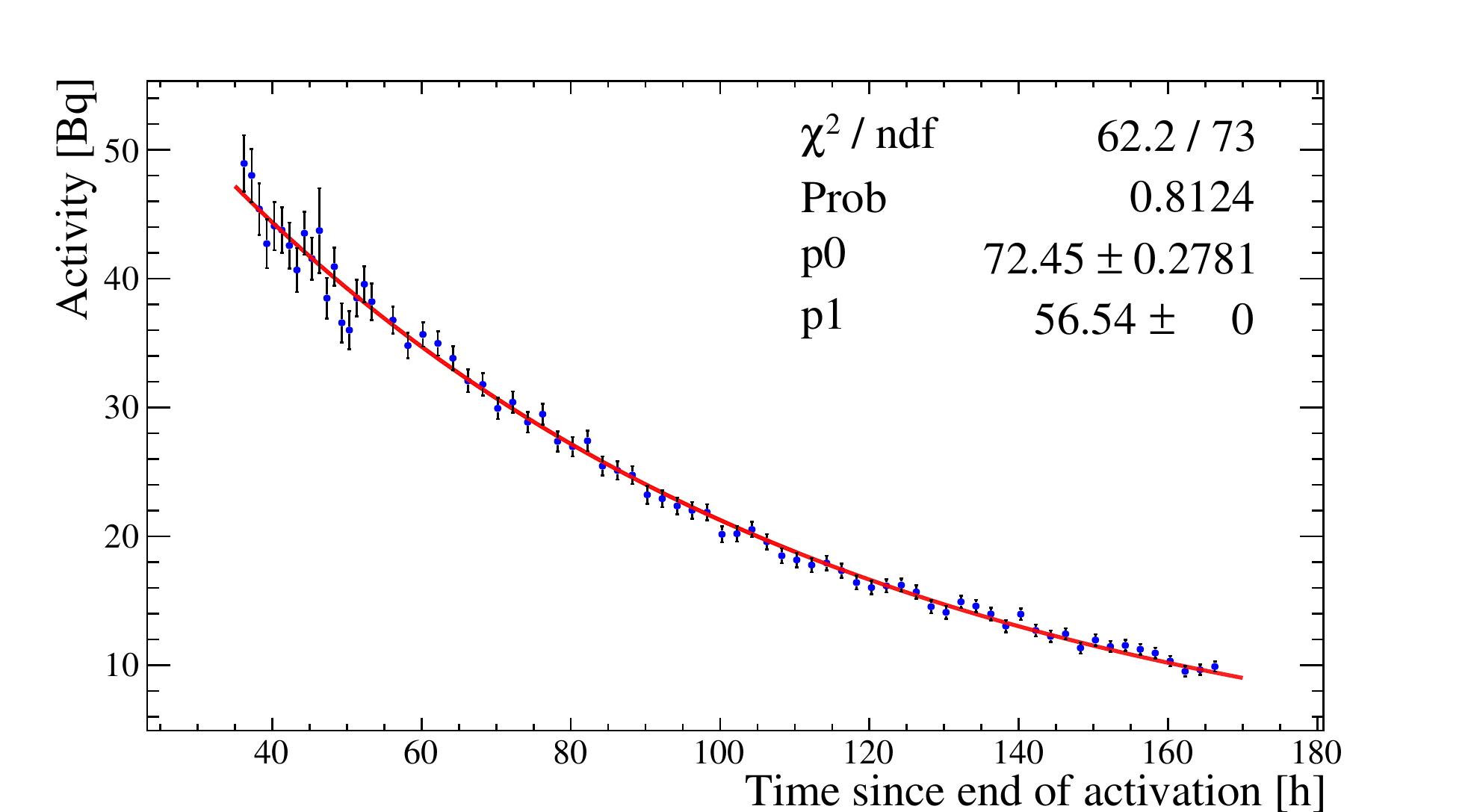}}
\caption{Time dependent \La[140] (left) and \Np[239] (right) activities, derived
from repeated analysis of gamma lines. The half-lives have been fixed to their tabulated values. The left panel shows data derived from interrupted counting of the same sample to allow counting of several samples on one detector. Repeated reinsertion of the sample results in an imperfect reproduction of the counting solid angle, visibly impacting the fit quality. The right panel shows data derived from uninterrupted counting.}
\label{fig:np239_act}
\end{figure}
The fit to the time dependence of the activities is compared to tabulated half-lives, further enhancing analysis selectivity. Once a nuclide assignment has been made the activity at some reference time is re-fit by fixing the half-life to its tabulated values as shown in Figure~\ref{fig:np239_act}. In this way all measurements contribute to a single activity value. For high statistics data, as shown in the left panel of Figure~\ref{fig:np239_act}, the data shows fluctuations beyond the statistical errors. These are due to small differences in the placement of the sample on the Ge detector and hence the counting solid angle, when the measurements of multiple samples necessitated the removal and later re-insertion of a particular sample. A point-by-point 10\% systematic error is added to the activities to account for this variability.

\section{Data and Conclusions}

EXO-200 has produced one of the lowest-background particle-interaction detectors ever built, in large part because of the exhaustive effort to screen construction materials for trace levels of radioactive contaminants.  We conclude our reporting of the results of these studies with many newly measured materials or, in some cases, re-confirmed measurements of new materials batches.  Results are reported in Table~\ref{TheTable}.   Combined with the previous work~\cite{rad1}, we have established an extensive library of material radio-purities suitable for use in planning the detailed design of new, next-generation detectors. 

\scriptsize
\clearpage
\onecolumn
\begin{landscape}
\begin{ThreePartTable}
\setlength{\LTcapwidth}{8.5in}
\begin{TableNotes}\footnotesize
\item[a] Out of secular equilibrium, result is from $^{228}$Ac only, $^{212}$Pb weak, $^{208}$Tl absent, possible effects of thin coating.
\item[b] Out of secular equilibrium, result is from $^{228}$Ac and $^{212}$Pb only, $^{208}$Tl weak, possible effects of thin coating.
\item[c] Out of secular equilibrium, result is from $^{208}$Tl only, $^{212}$Pb weak, $^{228}$Ac absent, possible effects of thin coating.
\end{TableNotes}
\begin{longtable}[t!]{p{.2in}p{4.0in}p{.6in}p{1.1in}p{1.1in}p{1.1in}}  
\caption{
Measurement results for \KTHU concentrations in a variety of materials.  Manufacturer production lot numbers or arbitrary batch identifiers  are indicated for materials where multiple lots were studied.  Limits are 90\%~C.L.  Uncertainties are quoted at 68\%~C.L. with a systematic error estimate included in quadrature, generally 10\% for NAA and Ge measurements and estimated by calibration data for ICPMS.  The GDMS measurement in entry~230 has a factor of two scaling uncertainty as shown.   In the ``method'' column, ``A.G. Ge'' refers to above-ground 
germanium counting.   Germanium results have been converted to elemental concentrations using average activities from multiple daughter nuclei. Entry numberings continue consecutively from Ref.~\cite{rad1}. Referenced entry numbers smaller than 226 can be found there.
}
\renewcommand*{\thefootnote}{\alph{footnote}}
\label{TheTable}
\\
Entry	&	Material 	&	Method	&\centering	K~conc. $[10^{-9} g/g]$ 		&\centering	Th~conc. $[10^{-12} g/g]$ 		&\centering	U~conc. $[10^{-12} g/g]$ 		\tabularnewline	
		\multicolumn{5}{c}{{\bf Lead}}												\tabularnewline	
226	&	JL Goslar re-smelted lead scrap from first EXO-200 lead production, batch 2638-1	&	ICP-MS	&\centering			&\centering	16.3$\pm$1.2		&\centering	14$\pm$9		\tabularnewline	
227	&	JL Goslar re-smelted lead scrap from first EXO-200 lead production, batch 2638-5	&	ICP-MS	&\centering			&\centering	16.5$\pm$1.2		&\centering	11.3$\pm$1.5		\tabularnewline	
228	&	JL Goslar re-smelted lead scrap from first EXO-200 lead production, batch 2640-1A	&	ICP-MS	&\centering			&\centering	12.4$\pm$1.4		&\centering	6$\pm$1		\tabularnewline	
229	&	JL Goslar re-smelted lead scrap from first EXO-200 lead production, batch 2638-5A	&	ICP-MS	&\centering			&\centering	17.2$\pm$1.8		&\centering	17.5$\pm$1.1		\tabularnewline	
230	&	Doe Run lead production lot D5031, Mayco Industries	&	GDMS	&\centering	11$^{+11}_{-6}$		&\centering	$<$10		&\centering	$<$9.0		\tabularnewline	
	&													\tabularnewline	
		\multicolumn{5}{c}{{\bf Plating}}												\tabularnewline	
231	&	.005\inch thick phosphor-bronze from entries~\idx{78} and~\idx{79}, acid cleaned by SLAC plating shop. No pre-analysis cleaning	&	ICP-MS	&\centering			&\centering	7.4$\pm$1.6		&\centering	6.8$\pm$1.4		\tabularnewline	
232	&	.005\inch thick phosphor-bronze from entries~\idx{78} and~\idx{79}, 1/2 Ni plated by SLAC plating shop. No pre-analysis cleaning	&	ICP-MS	&\centering			&\centering	8.5$\pm$1.7		&\centering	8.8$\pm$1.4		\tabularnewline	
233	&	.005\inch thick phosphor-bronze from entries~\idx{78} and~\idx{79}, Fully Ni plated by SLAC plating shop. No pre-analysis cleaning	&	ICP-MS	&\centering			&\centering	8.0$\pm$1.7		&\centering	9.3$\pm$1.5		\tabularnewline	
234	&	.01\inch thick phosphor-bronze from entry~\idx{80}, control sample for plating entries~\idx{339} through~\idx{340}	&	ICP-MS	&\centering			&\centering	2.4$\pm$0.5		&\centering	2.0$\pm$0.6		\tabularnewline	
235	&	.01\inch thick phosphor-bronze from entry~\idx{80}, plated with Pt over Ni by Technic Inc.	&	ICP-MS	&\centering			&\centering	22.3$\pm$0.9		&\centering	2.8$\pm$0.6		\tabularnewline	
236	&	.01\inch thick phosphor-bronze from entry~\idx{80}, plated with Ni by SLAC shop	&	ICP-MS	&\centering			&\centering	17.5$\pm$2.6		&\centering	12.2$\pm$1.3		\tabularnewline	
237	&	.01\inch thick phosphor-bronze from entry~\idx{80}, control sample for plating entries~\idx{342} through~\idx{340}	&	ICP-MS	&\centering			&\centering	$<$2.6		&\centering	5.7$\pm$1.0		\tabularnewline	
238	&	.01\inch thick phosphor-bronze from entry~\idx{80}, plated with Pt over Ni  by SLAC shop	&	ICP-MS	&\centering			&\centering	5.2$\pm$0.9		&\centering	8.1$\pm$1.4		\tabularnewline	
239	&	.005\inch thick phosphor-bronze from entry~\idx{78} and~\idx{79}, plated with Ni by SLAC shop	&	ICP-MS	&\centering			&\centering	26.2$\pm$1.4		&\centering	52.6$\pm$1.9		\tabularnewline	
240	&	.005\inch thick phosphor-bronze from entry~\idx{78} and~\idx{79}, plated with Ni +Au by SLAC shop	&	ICP-MS	&\centering			&\centering	25.6$\pm$2.7		&\centering	51.0$\pm$2.0		\tabularnewline	
241	&	.005\inch thick phosphor-bronze from entry~\idx{78} and~\idx{79}, plated with Ni +Rh by SLAC shop	&	ICP-MS	&\centering			&\centering	25.8$\pm$2.3		&\centering	51.4$\pm$2.7		\tabularnewline	
242	&	In-plating by SLAC shop on Phophor-Bronze, Material~from entry \idx{166} and ~\idx{167}, a spectator sample for gaskets produced simultaneously	&	ICP-MS	&\centering			&\centering	44.1$\pm$0.2		&\centering	31.7$\pm$1.9		\tabularnewline	
243	&	Same as entry~\idx{312}, but different plating run, several months later	&	ICP-MS	&\centering			&\centering	$<$1.2		&\centering	$<$1.3		\tabularnewline	
244	&	.01\inch thick phosphor bronze from entry~\idx{80}, control blank for entries~\idx{392} through~\idx{394}. No pre-analysis cleaning	&	ICP-MS	&\centering			&\centering	3.1$\pm$0.8		&\centering	$<$4.6		\tabularnewline	
245	&	Same as entry~\idx{391} with 1000 A Al + 500 A $\rm MgF_2$ plating on one side. No pre-analysis cleaning	&	ICP-MS	&\centering			&\centering	6.2$\pm$0.9		&\centering	$<$5.2		\tabularnewline	
246	&	Same as entry~\idx{391} with 1000 A Al + 500 A $\rm MgF_2$ plating on one side, 300A Ni +1000 A Au on other side. No pre-analysis cleaning	&	ICP-MS	&\centering			&\centering	9.0$\pm$1.5		&\centering	11.1$\pm$1.3		\tabularnewline	
247	&	Same as entry~\idx{391} with 300 A Ni + 1000 A Au on one side. No pre-analysis cleaning	&	ICP-MS	&\centering			&\centering	9.4$\pm$0.9		&\centering	10.9$\pm$1.3		\tabularnewline	
248	&	1/8\inch thick copper scraps from entry~\idx{5}, control sample for plating entries~\idx{336} through~\idx{337}	&	ICP-MS	&\centering			&\centering	4.6$\pm$1.6		&\centering	5.3$\pm$1.7		\tabularnewline	
249	&	1/8\inch thick copper, entry~\idx{5}, plated with Pt over Ni by Technic Inc.	&	ICP-MS	&\centering			&\centering	5.8$\pm$1.3		&\centering	6.8$\pm$1.3		\tabularnewline	
250	&	1/8\inch thick copper, entry~\idx{5}, plated with Pt over Ni by SLAC shop	&	ICP-MS	&\centering			&\centering	$<$8.3		&\centering	9.6$\pm$1.4		\tabularnewline	
251	&	Ni+Au plating by VPEI on copper from entries~\idx{1} and~\idx{2}, No pre-analysis cleaning	&	ICP-MS	&\centering			&\centering	$<$8.6		&\centering	13.7$\pm$2.1		\tabularnewline	
252	&	Al plating by VPEI on copper from entries~\idx{2} and~\idx{1}. No pre-analysis cleaning	&	ICP-MS	&\centering			&\centering	12.9$\pm$1.9		&\centering	17.2$\pm$2.9		\tabularnewline	
	&													\tabularnewline	
		\multicolumn{5}{c}{{\bf Flex Cables}}												\tabularnewline	
253	&	Coverlay, Dupont FR 70001, 13~$\mu$m polymide + 13~$\mu$m acrylic adhesive	&	ICP-MS	&\centering			&\centering	$<$1.6	 $\rm pg/cm^2$	&\centering	$<$3.6	 $\rm pg/cm^2$	\tabularnewline	
254	&	Coverlay, Dupont FR 0110, 25~$\mu$m polymide + 25~$\mu$m acrylic adhesive	&	ICP-MS	&\centering			&\centering	$<$1.8	 $\rm pg/cm^2$	&\centering	$<$5.4	 $\rm pg/cm^2$	\tabularnewline	
255	&	Coverlay, ShinEtsu MicroSi CA 333, 25~$\mu$m polymide + 25~$\mu$m epoxy adhesive	&	ICP-MS	&\centering			&\centering	$<$1.6	 $\rm pg/cm^2$	&\centering	34.2$\pm$2.8	 $\rm pg/cm^2$	\tabularnewline	
256	&	Coverlay, ShinEtsu MicroSi CA 335, 25~$\mu$m polymide + 35~$\mu$m epoxy adhesive	&	ICP-MS	&\centering			&\centering	72.5$\pm$2.6	 $\rm pg/cm^2$	&\centering	93$\pm$3	 $\rm pg/cm^2$	\tabularnewline	
257	&	Coverlay, ShinEtsu MicroSi CA 338	&	ICP-MS	&\centering			&\centering	54$\pm$7	 $\rm pg/cm^2$	&\centering	76$\pm$9	 $\rm pg/cm^2$	\tabularnewline	
258	&	Copper layer of flexible cables custom etched by FlexCTech, from material in entry~\idx{107}, cleaned with acetone and ethanol before digestion	&	ICP-MS	&\centering			&\centering	$<$2.3	 $\rm pg/cm^2$	&\centering	5.5$\pm$0.6	 $\rm pg/cm^2$	\tabularnewline	
259	&	Same as entry~\idx{479}, soaked in 50\% HCl for 30 seconds before analysis digestion	&	ICP-MS	&\centering			&\centering	$<$2.1	 $\rm pg/cm^2$	&\centering	7.5$\pm$0.7	 $\rm pg/cm^2$	\tabularnewline	
260	&	Same as entry~\idx{479}, soaked in 50\% HCl for 30 seconds and $\rm HNO_3$ for 10 seconds	&	ICP-MS	&\centering			&\centering	$<$1.7	 $\rm pg/cm^2$	&\centering	5.7$\pm$0.7	 $\rm pg/cm^2$	\tabularnewline	
261	&	Same as entry~\idx{479}, etched with $\rm N_2$, $\rm O_2$, and $\rm CF_4$ plasmas by Data Electronics	&	ICP-MS	&\centering			&\centering	$<$2.3	 $\rm pg/cm^2$	&\centering	4.7$\pm$0.7	 $\rm pg/cm^2$	\tabularnewline	
262	&	Kapton layer of flexible cables custom etched by FlexCTech, from material in entry~\idx{107}	&	ICP-MS	&\centering			&\centering	$<$2.3	 $\rm pg/cm^2$	&\centering	3.4$\pm$0.6	 $\rm pg/cm^2$	\tabularnewline	
263	&	Copper layer from material in entry~\idx{107}, control travelled with samples ~\idx{479} and~\idx{362}	&	ICP-MS	&\centering			&\centering	$<$2.3	 $\rm pg/cm^2$	&\centering	$<$1.4	 $\rm pg/cm^2$	\tabularnewline	
264	&	Kapton layer from material in entry~\idx{107}, control travelled with samples ~\idx{479} and ~\idx{362}	&	ICP-MS	&\centering			&\centering	$<$2.5	 $\rm pg/cm^2$	&\centering	$<$1.6	 $\rm pg/cm^2$	\tabularnewline	
	&													\tabularnewline	
		\multicolumn{5}{c}{{\bf Miscellaneous}}												\tabularnewline	
265	&	Cusil brazing material, WESGO Inc. GmBH	&	ICP-MS	&\centering			&\centering	16$\pm$1		&\centering	14.5$\pm$0.6		\tabularnewline	
266	&	Copper (entry~\idx{5}), e-beam welded at Applied Fusion Inc.	&	ICP-MS	&\centering			&\centering	9.7$\pm$2.5		&\centering	$<$15		\tabularnewline	
267	&	Screws machined from Material \idx{175}	&	ICP-MS	&\centering			&\centering	17$\pm$3		&\centering	83$\pm$9		\tabularnewline	
268	&	Metalization disolved from EXO-200 production APD's made with EXO selected metals.  Batch 1. Compare with stock APD, entry~\idx{127}	&	ICP-MS	&\centering	$<$3.1	 $\rm ng/APD$	&\centering	$<$2.1	 $\rm pg/APD$	&\centering	6.7$\pm$1.6	 $\rm pg/APD$	\tabularnewline	
269	&	EXO-200 APD metalization, batch 2	&	ICP-MS	&\centering	$<$4.8	 $\rm ng/APD$	&\centering	$<$2.7	 $\rm pg/APD$	&\centering	8.9$\pm$1.6	 $\rm pg/APD$	\tabularnewline	
270	&	EXO-200 APD metalization, batch 3	&	ICP-MS	&\centering	$<$5.3	 $\rm ng/APD$	&\centering	4.4$\pm$0.8	 $\rm pg/APD$	&\centering	$<$6.6	 $\rm pg/APD$	\tabularnewline	
271	&	Polyimide Optical Fiber, Ceramoptec Industries, Geometries: 50/125/-/140	&	NAA	&\centering	3\,100$\pm$300		&\centering	180$\pm$60		&\centering	$<$270		\tabularnewline	
272	&	Polyimide Optical Fiber, Ceramoptec Industries Inc.,  Geometries: 90/100/-/107	&	NAA	&\centering	1\,300$\pm$140		&\centering	$<$300		&\centering	$<$100		\tabularnewline	
273	&	Weldable high strength bronze rod, alloy 655, McMaster-Carr part \#89575K84	&	ICP-MS	&\centering			&\centering	51.1$\pm$2.4		&\centering	88$\pm$4		\tabularnewline	
274	&	TPC field cage resistors, sapphire subtrate (entry~\idx{155}), conductive paste (entry~\idx{158}), resistive paste (entry~\idx{157}), manufactured by Piconics, Tyngsboro, MA	&	ICP-MS	&\centering			&\centering	$<$6.9	 $\rm pg/ea$	&\centering	$<$7.0	 $\rm pg/ea$	\tabularnewline	
275	&	Gold for APD-plane coating, Cerac Inc.	&	ICP-MS	&\centering			&\centering	40.7$\pm$0.4		&\centering	$<$13		\tabularnewline	
276	&	Nickel for APD coating, Cerac Inc.	&	ICP-MS	&\centering			&\centering	15$\pm$4		&\centering	$<$15		\tabularnewline	
277	&	$\rm MgF_{2}$ powder for APD coating, Cerac Inc. 	&	Ge	&\centering	$<$1\,100		&\centering	3\,400$\pm$600$\,^a$		&\centering	2\,800$\pm$300\tnote[a]\footnotemark[1]\footnotetext[1]{Out of secular equilibrium, result is from $^{228}$Ac only, $^{212}$Pb weak, $^{208}$Tl absent, possible effects of thin coating.}		\tabularnewline	
278	&	VCR copper gaskets, 27 5/8\inch, (Swagelok CU-10-VCR-2)	&	ICP-MS	&\centering			&\centering	29.8$\pm$2.1		&\centering	40$\pm$2		\tabularnewline	
279	&	VCR copper gaskets, 29 1\inch, (Swagelok CU-16-VCR-2)	&	ICP-MS	&\centering			&\centering	25.6$\pm$2.4		&\centering	28.9$\pm$1.3		\tabularnewline	
280	&	Swagelok 3/8\inch brass unions for calibration tubing	&	A.G. Ge	&\centering	4\,400$\pm$2\,000		&\centering	1\,200$\pm$1\,000		&\centering	$<$4\,900		\tabularnewline	
281	&	Si-bronze flat washers from Bolt Depot, size 8, product \#3489	&	ICP-MS	&\centering			&\centering	88$\pm$5		&\centering	20$\pm$6		\tabularnewline	
282	&	Si-bronze flat washers from Bolt Depot, size 10, product \#3490	&	ICP-MS	&\centering			&\centering	74$\pm$4		&\centering	12$\pm$3		\tabularnewline	
283	&	Si-bronze machine screws from Bolt Depot, size 8-32 x 1-1/4\inch, product \#3427	&	ICP-MS	&\centering			&\centering	72$\pm$10		&\centering	$<$9.8		\tabularnewline	
284	&	Si-bronze machine screws from Bolt Depot, size 10-32 x 3/4\inch, product \#3440	&	ICP-MS	&\centering			&\centering	73$\pm$4		&\centering	15$\pm$4		\tabularnewline	
285	&	Si-bronze machine screws from Bolt Depot, size 10-32 x 1-1/4\inch, product \#3442	&	ICP-MS	&\centering			&\centering	46$\pm$7		&\centering	$<$9.9		\tabularnewline	
286	&	Teflon tubing, 3/16\inch ID 1/4\inch OD, McMaster-Carr	&	A.G. Ge	&\centering	$<$150		&\centering	$<$3\,600		&\centering	$<$18\,000		\tabularnewline	
287	&	PFA tubing, OD 1/8\inch, wall 0.030\inch, Swagelok PFA-T2-030-100, Grade: PFA AP-230, Trace ID: 78390	&	Ge	&\centering	$<$2\,700		&\centering	$<$1\,100		&\centering	$<$480		\tabularnewline	
288	&	Cu tubing, OD 3/16\inch, wall 0.8mm, Metallica SA	&	ICP-MS	&\centering			&\centering	12.1$\pm$2.6		&\centering	13$\pm$3		\tabularnewline	
289	&	Cu tubing, OD 3/16\inch, wall 0.8mm, Metallica SA	&	Ge	&\centering	$<$480		&\centering	$<$520		&\centering	$<$96		\tabularnewline	
290	&	Cu tubing, OD 1/4\inch, wall 0.8mm, Metallica SA	&	ICP-MS	&\centering			&\centering	8.2$\pm$1.5		&\centering	7$\pm$2		\tabularnewline	
291	&	Cu tubing, OD 1/4\inch, wall 0.8mm, Metallica SA	&	Ge	&\centering	$<$370		&\centering	$<$140		&\centering	$<$200		\tabularnewline	
292	&	Cu tubing, OD 3/8\inch, wall 0.8mm, Metallica SA	&	ICP-MS	&\centering			&\centering	5.0$\pm$1.1		&\centering	$<$6.1		\tabularnewline	
293	&	Cu tubing, OD 3/8\inch, wall 0.8mm, Metallica SA	&	Ge	&\centering	$<$300		&\centering	$<$280		&\centering	$<$76		\tabularnewline	
294	&	1mm ETP Cu wire, medium hard. Metallica SA, P/N: HM 2000 098	&	Ge	&\centering	$<$2\,700		&\centering	$<$1\,800		&\centering	$<$610		\tabularnewline	
295	&	1mm ETP Cu wire from Metallica SA, P/N: HM 2000 098	&	ICP-MS	&\centering			&\centering	45$\pm$3		&\centering	11$\pm$3		\tabularnewline	
296	&	Thermocouple wire,  Copper-Constantan, Omega Engineering, Inc.,  part \# TT-T-30-SLE  (ROHS)	&	Ge	&\centering	$<$1\,200		&\centering	$<$540		&\centering	$<$400		\tabularnewline	
297	&	Diamond coated polishing foil collection, McMaster Carr 8258A15 through 8258A19, tyically 7.5\~g/sheet	&	A.G. Ge	&\centering	$<$9\,700		&\centering	70\,000$\pm$16\,000		&\centering	18\,000$\pm$6\,000		\tabularnewline	
298	&	Tungsten wire from Goodfellow, $\rm 200~\mu m$ dia. 99.95\% purity	&	A.G. Ge	&\centering	$<$220\,000		&\centering	$<$460\,000		&\centering	$<$270\,000		\tabularnewline	
299	&	Stanford Research Systems RGA filament, ThO$_2$ coated	&	A.G. Ge	&\centering	15\,000\,000$\pm$3\,000\,000	 $\rm ng/filament$	&\centering	1\,900\,000\,000$\pm$500\,000\,000	 $\rm pg/filament$	&\centering	$<$1\,700\,000	 $\rm pg/filament$	\tabularnewline	
300	&	RGA filament,  YO$_2$ coated, E-filaments	&	Ge	&\centering	$<$130\,000	 $\rm ng/filament$	&\centering	10\,700\,000$\pm$1\,100\,000$\,^b$	 $\rm pg/filament$	&\centering	$<$37\,000\footnotemark[2]\footnotetext[2]{Out of secular equilibrium, result is from $^{228}$Ac and $^{212}$Pb only, $^{208}$Tl weak, possible effects of thin coating.}	 $\rm pg/filament$	\tabularnewline	
301	&	Stanford Research Systems RGA filament,  YO$_2$ coated, Fredericks	&	Ge	&\centering	$<$69\,000	 $\rm ng/filament$	&\centering	270\,000$\pm$80\,000$\,^a$	 $\rm pg/filament$	&\centering	$<$18\,000\footnotemark[1]	 $\rm pg/filament$	\tabularnewline	
302	&	Two component epoxy, MasterBond EP29LPSP, Mfr. Date July 5, 2006	&	Ge	&\centering	$<$1\,600		&\centering	$<$740		&\centering	760$\pm$180		\tabularnewline	
303	&	Two component epoxy, MasterBond EP29LPSP, Mfr. Date Feb 12, 2008	&	Ge	&\centering	$<$930		&\centering	$<$540		&\centering	$<$360		\tabularnewline	
304	&	Buna-N O-rings, Grotenrath Rubber Products Inc.	&	A.G. Ge	&\centering	27$\pm$4		&\centering	23\,400$\pm$1\,200		&\centering	2\,200$\pm$300		\tabularnewline	
305	&	Dow Corning high vacuum grease	&	Ge	&\centering	$<$1\,200		&\centering	$<$240		&\centering	$<$3\,800		\tabularnewline	
306	&	Silicone RTV sealant, NuSil CV-1142, Silicone Technology	&	Ge	&\centering	$<$3\,200		&\centering	5\,700$\pm$1\,600$\,^c$		&\centering	$<$2\,000\footnotemark[3]\footnotetext[3]{Out of secular equilibrium, result is from $^{208}$Tl only, $^{212}$Pb weak, $^{228}$Ac absent, possible effects of thin coating.}		\tabularnewline	
307	&	Vacseal high vacuum sealant, Prouct \#5052	&	A.G. Ge	&\centering	$<$12\,000		&\centering	$<$5\,700		&\centering	$<$6\,200		\tabularnewline	
308	&	APT teflon, 0.028\inch thick, from EXO-200, re-cleaned, control for entry~\idx{428}	&	NAA	&\centering	2.16$\pm$0.23		&\centering	$<$0.65		&\centering	$<$1.3		\tabularnewline	
309	&	APT teflon, 0.060\inch thick, from EXO-200, re-cleaned, control for entry~\idx{429}	&	NAA	&\centering	2.01$\pm$0.20		&\centering	$<$0.62		&\centering	$<$1.2		\tabularnewline	
310	&	APT teflon, 0.028\inch thick, as installed in EXO-200, interpreted as surface contamination	&	NAA	&\centering	3.4$\pm$0.3	 $\rm ng/cm^2$	&\centering	0.23$\pm$0.06	 $\rm pg/cm^2$	&\centering	$<$0.22	 $\rm pg/cm^2$	\tabularnewline	
311	&	APT teflon, 0.060\inch thick, as installed in EXO-200, interpreted as surface contamination	&	NAA	&\centering	12.0$\pm$1.2	 $\rm ng/cm^2$	&\centering	0.93$\pm$0.09	 $\rm pg/cm^2$	&\centering	0.19$\pm$0.06	 $\rm pg/cm^2$	\tabularnewline	
312	&	Polyethylene foam, McMaster-Carr P/N 86155K33, 5~cm thick, no cleaning	&	A.G. Ge	&\centering	$<$700\,000		&\centering	$<$2\,600\,000		&\centering	$<$290\,000		\tabularnewline	
313	&	Minwax water based Polycrylic protective finish (clear satin)	&	A.G. Ge	&\centering	1\,100\,000$\pm$26\,000		&\centering	$<$1\,500		&\centering	$<$4\,000		\tabularnewline	
314	&	Carbo-Act activated charcoal	&	A.G. Ge	&\centering	5\,900$\pm$400		&\centering	7\,600$\pm$800		&\centering	$<$6\,000		\tabularnewline	
315	&	Non-specific granular charcoal sample	&	A.G. Ge	&\centering	16\,730\,000$\pm$70\,000		&\centering	270\,000$\pm$10\,000		&\centering	72\,000$\pm$8\,000		\tabularnewline	
316	&	Granular activated carbon used in deradonator, Calgon Carbon Corp, OVC 4x8	&	A.G. Ge	&\centering	15\,500\,000$\pm$1\,600\,000		&\centering	250\,000$\pm$30\,000		&\centering	118\,000$\pm$15\,000		\tabularnewline	

\insertTableNotes
\end{longtable} 
\setlength{\LTcapwidth}{4in}
\end{ThreePartTable}
\end{landscape}
\normalsize
\clearpage

\begin{appendices}
\section{Reactor neutrons}
\label{sec:neutrons}
At nuclear reactors the neutron flux shows a broad distribution in Energy, $E$. It is composed of three components: thermal $\Phi_{th}(E,T)$, epi-thermal $\Phi_{et}(E)$ and fast fission $\Phi_{f}(E)$ neutrons. The total flux is modeled as the sum of these components: $\Phi(E,T)=\Phi_{th}(E,T)+\Phi_{et}(E)+\Phi_{f}(E)$. For a detailed discussion of the modeling of the first two flux components see Ref.~\cite{wescott_1955}. 

At low neutron energy the  flux is described by the Maxwell-Boltzmann distribution, 
$\Phi_{th}(E,T)$:
\begin{equation}
\Phi_{th}(E,T)\; dE\; =\; \sqrt{\frac{8}{\pi \cdot m}}\cdot \frac{n}{(k_B\cdot T)^{3/2}}\cdot 
E\cdot e^{-\frac{E}{k_B\cdot T}}\; dE, \label{eq:maxwell_flux_dist}
\end{equation}
where $m$ is the neutron mass, $n$ is the integral neutron density, $k_B$ is Boltzmann constant, and $T$ denotes the neutron temperature, assumed to be equal to the temperature of the water moderator.

Here the neutron flux is taken to be proportional to $v_n\cdot n_{th}(v_n,T)$, with $v_n$ denoting the neutron velocity and $n_{th}(v_n,T)$,
the Maxwell velocity distribution of the neutron {\it density}. 
The total thermal neutron flux, $\Phi_{th}$, is defined as: $\Phi_{th}\equiv \int_{0}^{\infty}\Phi_{th}(E,T)\; dE=\langle v_n \rangle \cdot n$, with $\langle v_n \rangle=\sqrt{\frac{8\cdot k_B\cdot T}{\pi \cdot m}}$ denoting the average neutron velocity, evaluated over the velocity distribution. 

Ref.~\cite{wescott_1955} states that for neutron energies larger than about $5\cdot k_B\cdot T$ the epi-thermal spectrum ``abruptly'' turns on. The epi-thermal flux, $\Phi_{et}(E)$, is known to be inversely proportional to the neutron kinetic energy: $\Phi_{et}(E)\; dE\; \; \propto \frac{dE}{E}$. When defining the total epi-thermal neutron flux $\Phi_{et}$ such that: 
\begin{displaymath}
\Phi_{et}\; =\; \int_{E_{l}^{et}}^{E_{u}^{et}}\Phi_{et}(E)\; dE, 
\end{displaymath}
one obtains the energy dependent epi-thermal neutron flux as:
\begin{equation}
\Phi_{et}(E)\; dE\; =\; \frac{\Phi_{et}}{\ln\left( \frac{E_{u}^{et}}{E_{l}^{et}} \right)}
\cdot \frac{dE}{E} \label{eq:et_flux}
\end{equation}
To arrive at a finite integrated flux, the epi-thermal flux component can only be defined over integration limits. 

At MeV-energies the reactor neutron spectrum is dominated by \U[235] fast fission neutrons. Because neutron capture cross sections are small at these high energies, including the fast fission flux component typically has a negligible effect on the analysis of NAA data. However, threshold reactions --- for example of $(n,p)$ type --- on the sample matrix itself can create large sample related backgrounds that limit the analysis sensitivity. Inclusion of this flux component is thus useful for the purpose of background estimation. Following Ref.~\cite{jendl4.0}, the energy distribution of \U[235] fast fission neutrons, $\Phi_f(E)$, can be modeled by the Watt spectrum: 
\begin{equation}
\Phi_f(E)\; dE = \Phi_f\cdot \left[\sqrt{\frac{4}{\pi\cdot a^3\cdot b}}\cdot e^{-\frac{a\cdot b}{4}}\right]\cdot e^{-\frac{E}{a}}
\cdot \sinh(\sqrt{b\cdot E})\; dE,
\label{eq:watt}
\end{equation}
with $a = 0.988$~MeV, and $b = 2.249$~MeV$^{-1}$. For an upper cut-off energy ($E^f_u$) of 10~MeV or larger (see Ref.~\cite{endf_manual} for a complete analytic expression for the spectral normalization, the term in square brackets in Equation~\ref{eq:watt}), $\Phi_f$ is the integrated fast neutron flux defined as: $\Phi_f=\int_{0}^{E^f_u}\Phi_f(E)\; dE$. At that chosen upper cut-off energy this approximate spectral normalization differs from the complete one by about 0.1\%. 

\subsection{Average cross sections}
The use of cross sections, averaged over the energy distribution of the three flux components discussed before, converts Equation~\ref{eq:daughter_decay_rate} into an algebraic equation. The JENDL-4.0 database~\cite{jendl4.0} conveniently provides such average cross sections, greatly simplifying the interpretation of NAA data. In the following we discuss how to use these tabulated cross sections. 

In JENDL-4.0 the thermal average cross section is defined as:
\begin{equation}
    \sigma _{th}(T_0) \; \equiv \; \frac{2}{\sqrt{\pi}}\cdot 
                                   \frac{\int_{E_{l}^{th}}^{E_{u}^{th}}\sigma(E)\cdot \Phi_{th}(E,T_0)\; dE}
                                   {\int_{E_{l}^{th}}^{E_{u}^{th}}\Phi_{th}(E,T_0)\; dE}\approx
                                   \frac{2}{\sqrt{\pi}}\cdot 
                                   \frac{Y_{th}}
                                   {\Phi_{th}}, \label{eq:average_sig_th}
\end{equation}
where $T_0=300$~K is the reference temperature. The lower and upper integration limits are $E_{l}^{th}=10^{-5}$~eV and $E_{u}^{th}=10$~eV, respectively. $Y_{th}$ is the thermal neutron induced reaction yield per atom. For the chosen integration limits the integral in the denominator of Equation~\ref{eq:average_sig_th} differs from $\Phi_{th}$ by order $10^{-6}$. The factor of $2/\sqrt{\pi}$ equals the ratio of the average over the most likely thermal neutron velocity for a Maxwell-distribution. It normalizes the total thermal neutron flux to the traditional notation which equates the integral flux to the volume density times the most likely neutron velocity. 

To model the capture of epi-thermal neutrons JENDL-4.0 provides the so-called  resonance integral (RI).  The RI is defined as: 
\begin{equation}
\sigma_{RI} \; \equiv \; \int_{E_{l}^{et}}^{E_{u}^{et}}\sigma(E)\; \frac{dE}{E}, \label{eq:ri}
\end{equation}
with $E_l^{et}=0.5$~eV and $E_u^{et}=10$~MeV. Because of the missing spectral-flux normalization, the tabulated RI is not an average cross section in a mathematical sense. Substituting Equation~\ref{eq:et_flux} into Equation~\ref{eq:ri} relates the RI to the epi-thermal neutron capture yield $Y_{et}$: 
\begin{equation}
        \sigma _{RI} \; =\; \ln \left( \frac{E_{u}^{et}}{E_{l}^{et}}\right) \cdot 
        \frac{\int_{E_{l}^{et}}^{E_{u}^{et}}\sigma(E)\cdot \Phi_{et}(E)\; dE}{\Phi_{et}}\; \approx \;
        16.81\cdot \frac{Y_{et}}{\Phi_{et}}\label{eq:ri_1}
\end{equation}
From Equation~\ref{eq:ri_1} one can see that a numerical factor is needed to relate the epi-thermal neutron capture yield to the resonance integral and epi-thermal flux integral. 

The average fast fission capture cross section, $\sigma_f$ is defined
as a proper mathematical average and relates directly to the capture yield $Y_f$ per
target atom:
\begin{equation}
      \sigma_f \; \equiv \; \frac{\int_{E_{l}^{f}}^{E_{u}^{f}}\sigma(E)\cdot \Phi_{f}(E)\; dE}
                            {\int_{E_{l}^{f}}^{E_{u}^{f}}\Phi_{f}(E)\; dE}\; \approx\;
                            \frac{Y_f}{\Phi_{f}},
\label{eq:fast_average}
\end{equation}
with $E_{u}^{f}=10^{-5}$~eV and $E_{u}^{f}=20$~MeV.

Using Equations~\ref{eq:average_sig_th},~\ref{eq:ri_1}, and~\ref{eq:fast_average}, Equation
~\ref{eq:daughter_decay_rate} can be re-written using the average cross sections and integral
neutron fluxes:
\begin{equation}
R_D(t)\; =\; N_T \cdot \left(1-e^{-t_i/\tau_D}\right)
\cdot e^{-t/\tau_D}\cdot \left[\sigma _{th}(T) \cdot \Phi_{th}+
\frac{\sigma _{RI}}{16.81}\cdot \Phi_{et}+
\sigma _{f} \cdot \Phi_{f} \right] \label{eq:daughter_decay_rate_1}
\end{equation}
$\Phi_{RI}=\frac{\Phi_{et}}{16.81}$ is called the resonance integral flux.
Numerical calculation of average cross sections done in the course of this work and using 
energy dependent neutron capture cross sections
for \Na[23], \K[41], \Cr[50], \Fe[58], \Co[59],
\Zn[64], \Au[197], \Th and \U,
obtained online from Brookhaven National Laboratory typically agreed to
within a few percent with the JENDL-4.0 tabulated values.

The water moderator at MITR II has a water temperature of about $T=340$~K, 11\% higher than the JENDL-4.0 reference temperature. For the nuclides listed before this results in a 6\% smaller thermal neutron capture rate compared to the JENDL-4.0 reference temperature, calculated using Equation~\ref{eq:average_sig_th}. The change corresponds to a factor of $\sqrt{\frac{T_0}{T}}$, as expected for thermal neutron capture cross sections which are proportional to $1/v_n$. The data is corrected for this temperature dependence.
 
\subsection{The reactor neutron flux}
\begin{figure}[htb!!!]\
\centering
\includegraphics[width=100mm]{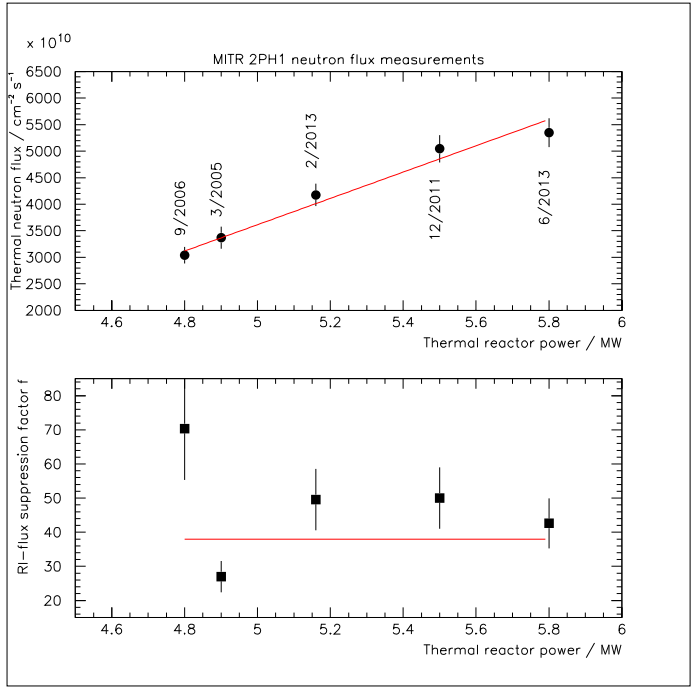}
\caption{\label{fig:n_flux}Thermal neutron flux (top) and ratio of RI-flux over thermal flux
(bottom) as determined from fly ash activations as a function of thermal reactor power. }
\end{figure}
Integral neutron fluxes are needed in Equation~\ref{eq:daughter_decay_rate_1} to determine $N_T$. The neutron flux at MITR's 2PH1 irradiation facility has been determined by activating NIST-certified fly ash (SRM 1633b), containing known amounts of activatable nuclides. Na, K, Ti, Cr, Fe, Th and U are used in a combined fit. The activities of all isotopes, $j$, are determined by Ge detector-based gamma-ray spectroscopy. Multiple gamma lines are used, whenever possible, in a time and energy differential fit, described in Sec.~\ref{sec:gamfit}. The parameterized values $A_j^P(t_d)$ of the activities at the reference time $t_d$ after end of irradiation are:  
\begin{displaymath}
A_j^P(t_d)= N_{T_j} \cdot \left(1-e^{-t_i/\tau_{D_j}}\right)
\cdot e^{-t_d/\tau_{D_j}}\cdot \left[ \sigma _{th}(T) \cdot \Phi_{th}+
\sigma _{RI}\cdot \Phi_{RI}+\sigma _{f} \cdot \Phi_{f} \right]\,,
\end{displaymath}

and the integral neutron fluxes are determined by minimizing:
\begin{displaymath}
\chi^2\; =\; \sum_{j=1}^7 \frac{\left[A_j^O(t_d)-A_j^P(t_d,t_i,\Phi_{th},\Phi_{RI},\Phi_f)
\right]^2}{\sigma_j^2}
\end{displaymath}
with respect to the three neutron fluxes, for the observed activities $A_j^O$.  The dependence of $A_j^P(t_d)$ on the irradiation parameters has now been made explicit, and the fluxes $\Phi_{th}$, $\Phi_{RI}$ and $\Phi_{f}$ are the free floating fit parameters.  The number of target atoms and their errors, $N_{T_j}$ and $\sigma_{N_{T_j}}$ are derived from the NIST certified elemental masses of the fly ash components and their errors. Molar masses and isotope abundances are taken from~\cite{table_of_isotopes}, the nuclide life times $\tau_{D_j}$ from~\cite{nndc}. These parameters have negligible uncertainties; their errors are not propagated.
Practically the determination of $\Phi_{RI}$ is driven by \Th and \U due to their large resonance integrals. \Ti[48]\unskip($n,p$)\Sc[48] determines $\Phi_{f}$. Due to this decoupling, the one-simga errors on the fluxes are determined as one-parameter errors, corresponding to $\Delta \chi^2=1$. The variances $\sigma_j^2$ contain the statistical counting error, a 10\% systematic added in quadrature to activities (meant to measure variability of the counting solid angle), and the error of the masses of the components of the fly ash, as certified by NIST. 

The thermal and epi-thermal neutron fluxes at MITR have been determined multiple times using the fly ash method. Figure~\ref{fig:n_flux} shows $\Phi_{th}$ (top) and $f=\Phi_{RI}/\Phi_{th}$ (bottom) as a function of thermal reactor power. These data were collected between 2005 and 2013. The data shown in the top panel of Figure~\ref{fig:n_flux} indicates a linear correlation between thermal neutron flux and reactor power, as long as the power is above a 4.8~MW threshold. Data collected at lower reactor power does not follow this systematic. However, for the data collected here this is irrelevant as these activations had their own flux calibrations and did not rely on the flux-power fit. A linear fit to the 2PH1 data above 4.8~MW yields a slope of $\rm (2.48\pm 0.27) \cdot 10^{13}$~cm$^{-2}$s$^{-1}$MW$^{-1}$, an intercept of $\rm (-8.8\pm 1.4)\cdot 10^{13}$~cm$^{-2}$s$^{-1}$ with an error correlation coefficient of -0.998. The average RI-flux suppression factor is found to be $\rm 38.0\pm 3.2$. 

The fast neutron flux has only been measured twice at 2PH1; a correlation with the reactor power cannot yet be established. For the activation at the highest neutron flux (6/2013) we determined: $\Phi_f=(3.3\pm 0.9)\cdot 10^{12}$~cm$^{-2}$s$^{-1}$ from fly ash data. In addition a sample of TiN was activated, constraining $\Phi_f$ via the reactions \Ti[46]\unskip($n$,$p$)\Sc[46] and \Ti[48]\unskip($n$,$p$)\Sc[48]. This independent measurement yielded $\Phi_f=(2.4\pm 0.2)\cdot 10^{12}$~cm$^{-2}$s$^{-1}$, averaged over both Sc-activities, and is in agreement with the fly ash data.

\end{appendices}

\ack
EXO-200 is supported by DOE and NSF in the United States,
NSERC in Canada, 
SNF in Switzerland, 
IBS in Korea,
RFBR in Russia,
DFG Cluster of Excellence ``Universe'' in Germany,
and CAS and ISTCP in China.

\bibliographystyle{elsart-num}
\bibliography{EXO-200-radiopurity-2017}






\end{document}